\documentclass[11pt]{article}

\usepackage[totalwidth=480pt, totalheight=680pt]{geometry}

\usepackage{vmargin}
\setmarginsrb{3 cm}{2.5 cm}{3 cm}{1.5 cm}{0 cm}{1 cm}{0 cm}{1 cm}

\usepackage{graphicx}

\usepackage{url}
\usepackage{bm}
\usepackage{enumitem}
\usepackage{xcolor}
\usepackage{booktabs}
\usepackage{hyperref}

\usepackage{xspace}
\usepackage{color}
\usepackage{algorithm,algorithmicx}
\usepackage[noend]{algpseudocode}
\usepackage{hyperref}
\usepackage{caption}
\usepackage{listings}
\usepackage{courier}
\usepackage{amsthm}
\usepackage{ulem}

\usepackage{amsmath}
\usepackage{parskip}
\usepackage{fancyhdr}
\usepackage{setspace,lipsum}

\title{\framework: A Framework for Facilitating Accurate and Interpretable Analytics for High Stakes Applications}
\author{K. Zheng et al.}
\date{Mar 19, 2020}

\makeatletter
\let\thetitle\@title
\let\theauthor\@author
\let\thedate\@date
\makeatother

\newcommand{\squishlist}
{
	\begin{list}{$\bullet$}
		{
			\setlength{\itemsep}{0pt}
			\setlength{\parsep}{3pt}
			\setlength{\topsep}{3pt}
			\setlength{\partopsep}{0pt}
			\setlength{\leftmargin}{1.5em}
			\setlength{\labelwidth}{1em}
			\setlength{\labelsep}{0.5em}
		}
	}
	
	\newcommand{\squishend}
	{
	\end{list}
}

\newcommand{\revise}[1]{{\color[HTML]{000000}#1}}
\newcommand{\highlight}[1]{\textit{#1}}
\newcommand{\powerpoint}[1]{\textbf{#1}}

\newcommand{\model}{\texttt{TITV}\xspace}
\newcommand{\framework}{\texttt{TRACER}\xspace}
\newcommand{\film}{\texttt{FiLM}\xspace}

\newcommand{\NEUP}{\texttt{NEUP}\xspace}
\newcommand{\ICAP}{\texttt{ICAP}\xspace}
\newcommand{\NP}{\texttt{NP}\xspace}
\newcommand{\WBC}{\texttt{WBC}\xspace}
\newcommand{\CarbonDioxide}{\texttt{CO2}\xspace}

\newcommand{\Oxygen}{\texttt{O2}\xspace}
\newcommand{\PH}{\texttt{PH}\xspace}
\newcommand{\BE}{\texttt{BE}\xspace}
\newcommand{\PCarbonDioxide}{\texttt{pCO2}\xspace}

\newcommand{\CRP}{\texttt{CRP}\xspace}
\newcommand{\NEU}{\texttt{NEU}\xspace}
\newcommand{\K}{\texttt{K}\xspace}
\newcommand{\NA}{\texttt{NA}\xspace}
\newcommand{\PTH}{\texttt{PTH}\xspace}
\newcommand{\URBC}{\texttt{URBC}\xspace}

\newcommand{\TEMP}{\texttt{TEMP}\xspace}
\newcommand{\MCHC}{\texttt{MCHC}\xspace}
\newcommand{\CP}{\texttt{CP}\xspace}
\newcommand{\AU}{\texttt{AU}\xspace}

\newcommand{\rev}[1]{{\color[HTML]{000000}#1}}

\newcommand\blfootnote[1]{%
	\begingroup
	\renewcommand{\thefootnote}{}
	\footnotetext{#1}
	\endgroup
}

\begin{document}

{\centering
{\Large \bfseries \thetitle}\\

\vspace{5mm}
{\large Kaiping Zheng$^\dag$, Shaofeng Cai$^\dag$, Horng Ruey Chua$^\S$,  Wei Wang$^\dag$, \\ Kee Yuan Ngiam$^\S$, Beng Chin Ooi$^\dag$} \\

\vspace{5mm}
$^\dag$National University of Singapore \hspace{15mm}
$^\S$National University Health System\\

{\scriptsize\{kaiping, shaofeng, wangwei, ooibc\}@comp.nus.edu.sg} \hspace{7mm}
{\scriptsize\{horng\_ruey\_chua, kee\_yuan\_ngiam\}@nuhs.edu.sg}
}

\pagestyle{fancy}
\vspace{5mm}
\begin{abstract}
\rev{In high stakes applications such as healthcare and finance analytics, the interpretability of predictive models is required and necessary for domain practitioners to trust the predictions.}
Traditional machine learning models, e.g., logistic regression (LR), are easy to interpret in nature.
However, many of these models aggregate time-series data without considering the temporal correlations and variations.
Therefore, their performance cannot match up to recurrent neural network (RNN) based models, which are nonetheless difficult to interpret.
In this paper, we propose a general framework \framework to \rev{facilitate accurate and interpretable predictions, with a novel model \model
devised for healthcare analytics and other high stakes applications such as financial investment and risk management.}
Different from LR and other existing RNN-based models, \model is designed to capture both the time-invariant and the time-variant feature importance
using a feature-wise transformation subnetwork and a self-attention subnetwork, for the feature influence shared over the entire time series and the time-related importance respectively.
\rev{
Healthcare analytics is adopted as a driving use case, and we note that the proposed \framework is also applicable to other domains, e.g., fintech.
We evaluate the accuracy of \framework extensively in two real-world hospital datasets, and our doctors/clinicians further validate the interpretability of \framework in both the patient level and the feature level.
Besides, \framework is also validated in a high stakes financial application and a critical temperature forecasting application.
The experimental results confirm that \framework facilitates both accurate and interpretable analytics for high stakes applications.}
\end{abstract}

\section{Introduction}
\label{sec:introduction}
\blfootnote{A version of this preprint will appear in ACM SIGMOD 2020.}
Database management systems (DBMS) have been extensively deployed to support OLAP-style analytics.
In modern applications, an ever-increasing number of data-driven machine learning based analytics have been developed with the support of DBMS for complex analysis~\cite{Boehm2016systemml, Zhang2014materialization, Zhang2010io, Kumar2015model}.
Particularly, there have been growing demands of machine learning for complex \rev{high stakes applications, such as healthcare analytics, financial investment, etc.}
\rev{Notably, healthcare analytics is a very important and complex application, which entails} data analytics on a selected cohort of patients for tasks such as diagnosis, prognosis, etc.
Various healthcare analytic models 
have been proposed for Electronic Medical Records (EMR) that record visits of patients to the hospital with time-series medical features.
In particular, neural network (NN) based models~\cite{choi2016doctor, lipton2015learning} have been shown to improve the accuracy over traditional machine learning models significantly. 
An accurate analytic model can help healthcare practitioners to make effective and responsive decisions on patient management and resource allocation.

\rev{However, accuracy alone is far from satisfactory for healthcare analytics and other high stakes applications in deployment. 
In this paper, we shall use healthcare analytics as our driving use case. Suppose an accurate model has been trained and deployed for in-hospital mortality prediction.}
Simply reporting to the doctors that 
``for this patient, the probability of mortality estimated by the model is $26\%$'' is unacceptable since in life-and-death medical decisions, a single number without proper explanations is meaningless for doctors to take interventions.
To alleviate this issue, it is essential to devise \powerpoint{an interpretable model}, explaining ``why''  certain decisions are made~\cite{molnar2018interpretable}, e.g., why a $26\%$ probability of mortality is produced for a specific patient. 
Such an interpretable model is critical to \rev{provide medically meaningful results to the doctors and insightful advice to the practitioners of other high stakes applications as well}.

Recently, various approaches have been proposed to explain the prediction results of neural network models~\cite{koh2017understanding}, \rev{some of which particularly focus on healthcare analytics~\cite{ma2017dipole, choi2016retain, sha2017interpretable}}. For example, attention modeling has been adopted to compute the \powerpoint{visit importance}, i.e., the importance of each visit of a patient~\cite{ma2017dipole} on the final prediction. 
Nonetheless, the visit importance only represents which visit is important and thus is not adequate to explain how and why the prediction is generated from the ``important'' visits.
Typically, doctors consider one visit to be more important when certain important indicators, i.e., medical features, of that visit deviate far from a normal range.
Therefore, the \powerpoint{feature importance} is more informative than the visit importance, and can be exploited to interpret model predictions.

However, the feature importance modeled in existing work is highly dependent on certain time periods~\cite{choi2016retain,  sha2017interpretable} without differentiating the \powerpoint{``time-invariant''} and \powerpoint{``time-variant''} feature importance.
\rev{We note that the medical feature, on the one hand, has a influence on a patient shared over the entire time series, and this is captured in the time-invariant feature importance; 
on the other hand, the influence of the feature on a patient can also vary over time periods or visits, and this is captured in the time-variant feature importance.}

\begin{figure}
\centering
\includegraphics[width=0.85\linewidth]{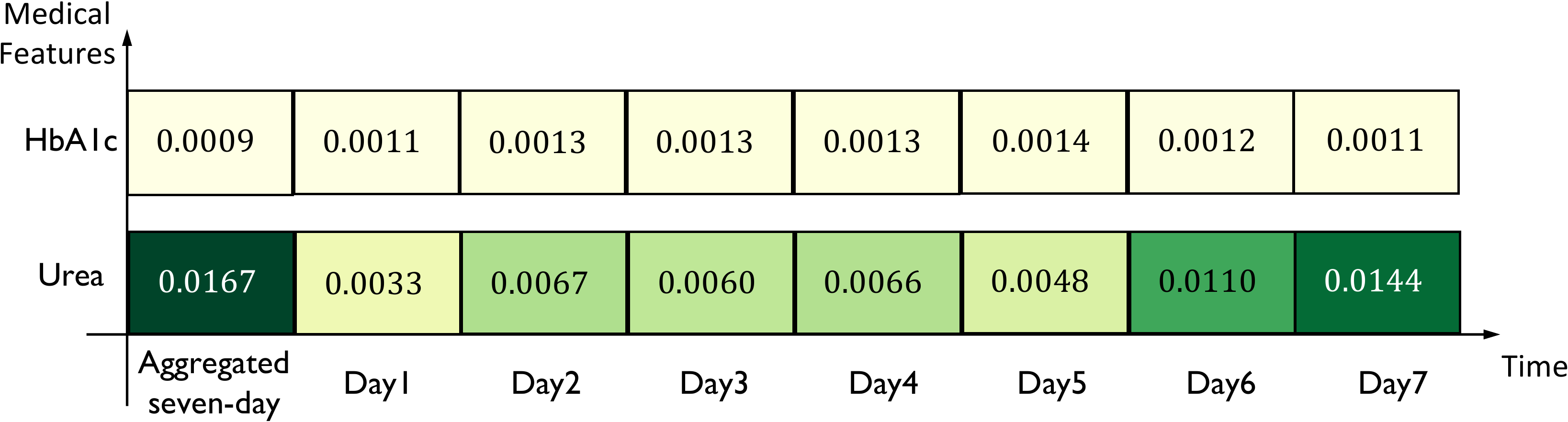}
\caption{ 
The normalized coefficients in \revise{both an LR model trained on the aggregated seven-day data (leftmost) and seven LR models trained separately}. We illustrate with two representative laboratory tests HbA1c and Urea (best viewed in color).}
\label{fig:example of local and global}
\end{figure}

For instance, we examine two representative laboratory tests, i.e., Glycated Hemoglobin (HbA1c) and Blood Urea Nitrogen (Urea), to predict the risk of patients developing acute kidney injury (AKI).
We first train one single logistic regression (LR) model by aggregating each medical feature across all the seven-day data (see the NUH-AKI dataset in Section~\ref{subsubsec:datasets and applications}).
The weight coefficient thus denotes the time-invariant feature importance, which reflects the influence of the corresponding feature on the final prediction. 
We also train seven LR models independently for the data of each day.
The learned coefficients of each day are regarded as the time-variant feature importance. 
\rev{
As shown in Figure~\ref{fig:example of local and global}\footnote{In each LR model (either time-invariant or time-variant), after training the model and obtaining the coefficients, we normalize the coefficients of all features via a Softmax function for illustration.},
we can see that HbA1c and Urea have different time-invariant feature importances which has been confirmed by doctors that Urea is more important in diagnosing kidney diseases~\cite{Urea}.
Specifically, the time-variant feature importance increases over days in Urea while it fluctuates in HbA1c.
The explanation from our doctors is that Urea is a key indicator of kidney dysfunction, so its importance grows approaching the AKI prediction time.
The same phenomenon can be found in Estimated Glomerular Filtration Rate (eGFR)~\cite{egfr},
which serves as a measure of assessing kidney function.
In contrast, HbA1c~\cite{hba1c}
is typically used to assess the risk of developing diabetes; hence, it has relatively low time-invariant feature importance and stable time-variant feature importance in the prediction.
}

\rev{To produce meaningful interpretation in healthcare analytics and other high stakes applications, we propose a general framework \framework to provide accurate and in\textbf{\underline{T}}erp\textbf{\underline{R}}et\textbf{\underline{A}}ble \textbf{\underline{C}}linical d\textbf{\underline{E}}cision suppo\textbf{\underline{R}}t to doctors and practitioners of other domains. Specifically, we devise a novel model \model for \framework, which captures both \textbf{\underline{T}}ime-\textbf{\underline{I}}nvariant and \textbf{\underline{T}}ime-\textbf{\underline{V}}ariant feature importance for each sample (e.g., each patient in healthcare analytics) in two different subnetworks,
with the former shared across time and the latter varying across different time periods.}
We adopt the feature-wise transformation mechanism, which is known as feature-wise linear modulation (\film)~\cite{dumoulin2018feature-wise, perez2018film, kim2017dynamic} in one subnetwork to model the time-invariant feature importance. 
While for the time-variant feature importance, we adopt a self-attention mechanism in the other subnetwork to model the fine-grained information. 
\revise{Trained with these two subnetworks of \model jointly in an end-to-end manner, \framework can produce accurate predictions \rev{and meanwhile, meaningful explanations in
analytics for high stakes applications.}}

The main contributions can be summarized as follows:
\begin{itemize}[leftmargin=*]
    \setlength\itemsep{0mm}
    \item We identify the research gaps in modeling the feature importance, where existing methods are overly dependent on certain time periods without differentiating the time-invariant and the time-variant feature importance.
    \rev{Contrarily, we demonstrate that both time-invariant and time-variant feature importance are essential for high stakes applications such as healthcare analytics}.

    \item We propose \revise{a general framework \framework to provide both accurate and interpretable \rev{decision support to doctors in healthcare analytics, and practitioners in other high stakes applications.} Specifically, in \framework, we devise a novel model \model that takes into consideration the time-invariant feature importance via the \film mechanism and the time-variant feature importance via the self-attention mechanism.}

    \item We evaluate the effectiveness of \framework for healthcare analytics in two real-world hospital datasets.
    Experimental results confirm that \framework can produce more accurate prediction results than state-of-the-art baselines.
    Meanwhile, both patient-level and feature-level interpretation results based on the feature importance of \framework have been validated by doctors to be medically meaningful,
    which shows that \framework can assist doctors in clinical decision making.
    
    \item \rev{We illustrate that \framework is also applicable to other high stakes applications.
    Specifically, \framework is further validated in a financial investment application and a temperature forecasting application.
    }
\end{itemize}

Our proposed \framework is integrated into GEMINI~\cite{ling2014gemini}, an end-to-end healthcare data analytics system, to facilitate the accuracy and interpretability of healthcare analytics.
The remainder of this paper is organized as follows.
In Section~\ref{sec:related work}, we review the related work. 
\revise{In Section~\ref{sec:framework}, we introduce our general \rev{framework \framework in the context of healthcare analytics}.
In Section~\ref{sec:methodology}, we elaborate the detailed design of the core component of \framework, the \model model.}
In Section~\ref{sec:experiments}, we analyze the experimental results \rev{of \framework in healthcare analytics, which demonstrate its effectiveness} in terms of both prediction accuracy and interpretation capability.
\rev{To illustrate that \framework is also applicable to other high stakes applications, we conduct experiments on a financial application and a temperature forecasting application in this section.}
We conclude in Section~\ref{sec:conclusion}.

\section{Related Work}
\label{sec:related work}

\revise{\subsection{Healthcare Analytics}}
\label{subsec:healthcare analytics}

\revise{Healthcare analytics is to conduct analytic tasks on patient data, which typically include diagnosis, prognosis, etc. Due to the recent advancements of DBMS~\cite{tan2019choosing, ratner2019sysml}, researchers manage to achieve more optimized support for healthcare analytics in terms of both effectiveness and efficiency~\cite{cao12smile}. Through high-quality healthcare analytics, we can provide medical professionals with useful insights on both diseases and patients, hence contributing to better patient management and faster medical research advancement.}

In recent years, EMR data collected from various hospitals and health organizations, becomes one of the major data sources for healthcare analytics. 
EMR data is heterogeneous in nature, consisting of information that ranges from patients' social and demographic information, to structured medical features such as diagnoses, laboratory tests, and further to unstructured data such as magnetic resonance images and doctors' notes. 
\rev{
The structured data can be fed into \framework directly, and the unstructured data can be converted into structured features before being input to \framework for analytics.
}

\begin{figure}
\centering
\includegraphics[width=0.6\linewidth]{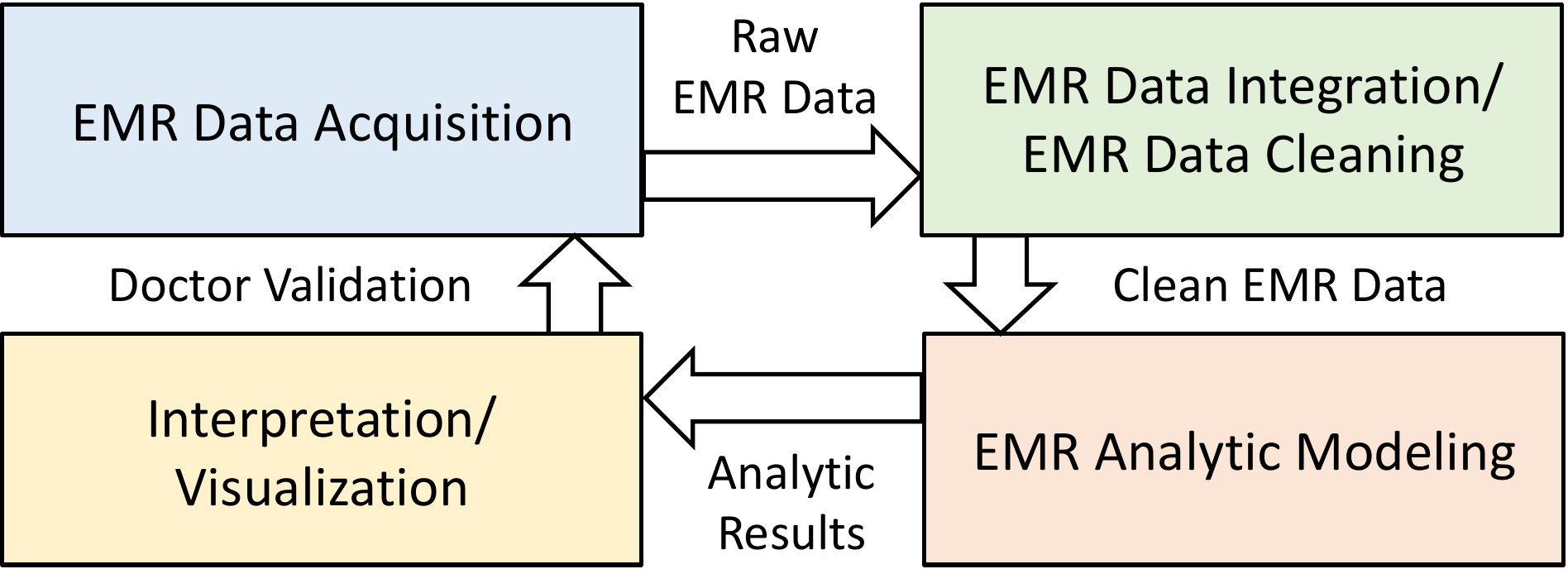}
\caption{EMR data analytics pipeline.}
\label{fig:healthcare analytics pipeline}
\end{figure}

The typical EMR data analytics pipeline from EMR Data Acquisition to Interpretation/Visualization is illustrated in Figure~\ref{fig:healthcare analytics pipeline}, where each module plays a critical role. 
To begin with, 
we collect raw EMR data, which may be quite ``dirty'' and thus may not be suitable for analytics directly.
Therefore, we need to feed the raw EMR data into the EMR Data Integration and EMR Data Cleaning module to improve the data quality.
Then the clean EMR data goes through the EMR Analytic Modeling module for analytics
and the analytic results are interpreted and visualized to render the derived medical insights easy to comprehend for medical experts, researchers, patients, etc. Finally, we make use of doctors' validation and feedback given based on the interpretation or visualization results to improve the whole EMR data analytics pipeline.
This EMR data analytics pipeline is supported by GEMINI~\cite{ling2014gemini}, a generalizable medical information analysis and integration platform, with the objective to design and implement an integrative healthcare analytic system in order to address various kinds of healthcare problems. GEMINI supports different key functionalities, such as capturing the feature-level irregularity in EMR data~\cite{zheng2017irregularity}, resolving the bias in EMR data~\cite{zheng2017bias} and an adaptive regularization method~\cite{luo2018regularization}. Our proposed \framework is integrated into GEMINI to facilitate the accuracy and interpretability of healthcare analytics.

\newpage
\revise{\subsection{Interpretability}}
\label{subsec:analytic result interpretation}
\revise{Interpretability, measuring the degree that the cause of a decision can be understood by human~\cite{biran2017explanation, miller2018explanation}, has drawn great attention in recent years.
When designing analytic models, we are faced with a trade-off between \highlight{what} is predicted (e.g., if a hospitalized patient will develop AKI in two days and what is the corresponding probability) and \highlight{why} a decision is made (e.g., which features indicate this patient's future development of AKI)~\cite{molnar2018interpretable}. 
The latter one which answers the \highlight{why} question lays emphasis on interpretability.
In some high stakes applications with high complexity and demanding requirements, it is not adequate to only know the answer to the \highlight{what} question, because the answer to the \highlight{why} question can improve the understanding of human on the problem, make human aware of when the model succeeds/fails, and boost the trust of human on the analytic results. Hence, it is vitally essential to take interpretability into consideration when designing analytic models.}

\vspace{2mm}
\revise{\subsection{Interpretability in Healthcare}}
\label{subsec:interpretability in healthcare analytics}

There exist some traditional machine learning methods which can achieve high interpretability~\cite{kho2011use, sun2012supervised, zhou2011multi, zhou2012modeling}; however, due to not considering the longitudinal property of EMR data, their performance may be degraded.

Deep learning~\cite{lecun2015deep} has recently aroused wide interest due to its ability to achieve state-of-the-art performance in a large number of applications~\cite{krizhevsky2012imagenet, sainath2013deep, mikolov2010recurrent}. In deep learning, there is a category of models, i.e., recurrent neural networks (RNN), such as the long short-term memory (LSTM) model~\cite{hochreiter1997long} and the gated recurrent unit (GRU) model~\cite{cho2014learning}, proposed to capture the dynamic behavior in sequential data. Although effective in modeling time-series data, the lack of interpretability tends to become an obstacle for deploying RNN-based models in healthcare analytics, despite some existing attempts~\cite{karpathy2015visualizing, che2015distilling}.
Fortunately, attention mechanism~\cite{bahdanau2014neural} comes into being, with which we can visualize the attention weights to facilitate the interpretability of healthcare analytics.

Some researchers have applied the attention mechanism in healthcare analytics~\cite{choi2017gram, bai2018interpretable}; nonetheless, these studies devise attention-based approaches for incorporating domain knowledge to improve prediction accuracy, rather than contributing to the interpretability of analytic results. 

Some other researchers employ the attention mechanism to facilitate interpretability. 
In~\cite{ma2017dipole}, an attention-based bidirectional RNN model is proposed to provide the interpretation for each visit in diagnosis prediction, but the more fine-grained interpretation for each medical feature is not supported.
In~\cite{choi2016retain}, RETAIN is proposed to facilitate interpretability via both visit-level attention and feature-level attention. However, these two levels of attention are connected by multiplication and the feature-level attention already conveys some information of the visit-level attention. 
Then in~\cite{sha2017interpretable}, GRNN-HA, a model with a hierarchical two-level attention mechanism is employed, but its two levels of attention for visits and for features are not connected.
\revise{Furthermore, both RETAIN and GRNN-HA only capture the time-variant feature importance which is heavily dependent on a certain visit, instead of a patient's whole time series.} 

\revise{Different from existing work, we highlight the necessity of capturing both the time-invariant and the time-variant feature importance in healthcare analytics at the same time.
In our proposed framework \framework, specifically, in the \model model, we combine the time-invariant feature importance via a \film mechanism and the time-variant feature importance via a self-attention mechanism to achieve this goal.}

\vspace{1mm}
\section{\revise{{\Large\framework} Framework}}
\label{sec:framework}

\rev{We propose a general framework \framework to facilitate accurate and interpretable decision support for healthcare analytics and other high stakes applications. In this section, we take healthcare analytics as an illuminating example to illustrate the architecture overview of \framework.}

\begin{figure}[t]
	\centering
	\includegraphics[width=0.7\linewidth]{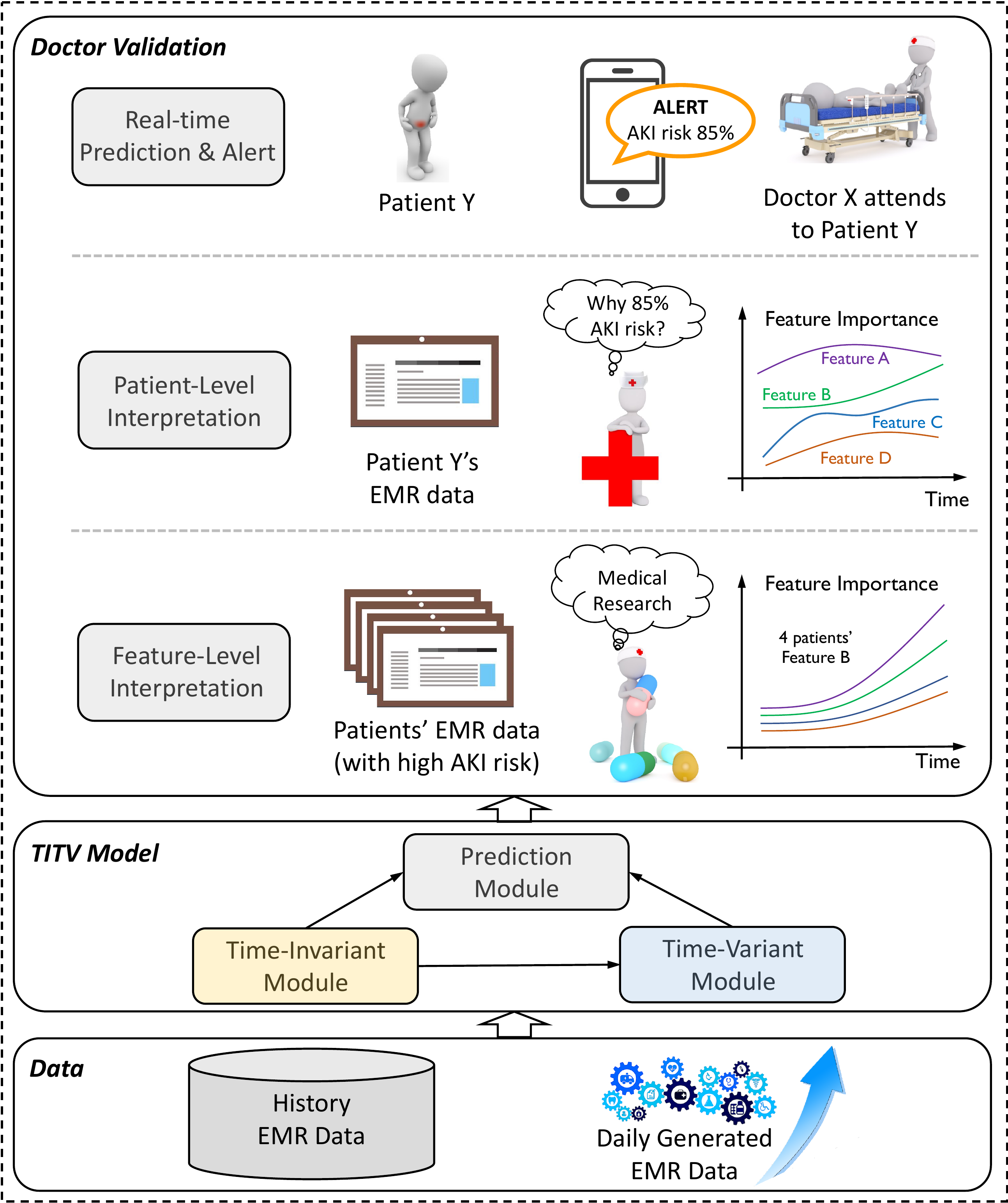}
	\caption{Overview of \framework in healthcare analytics.
	}
	\label{fig:framework}
\end{figure}

As shown in Figure~\ref{fig:framework}, \framework makes use of both the history time-series EMR data and the daily generated EMR data, and then feeds these data to the core component \model of \framework for analytics, in which both the time-invariant and the time-variant feature importance are captured in modeling.
Based on the analytic results, \framework supports doctor validation
with accurate and interpretable clinical decisions in scenarios including real-time prediction \& alert, patient-level interpretation and feature-level interpretation.

\vspace{2mm}
\noindent
\revise{\powerpoint{Data.} \framework incorporates two data sources: (i) the history time-series EMR data stored in the EMR database system of hospitals to obtain satisfactory performance of the analytic models; (ii) the daily generated EMR data (e.g., the data collected from a new patient to the hospital, or some new laboratory tests measured for a hospitalized patient) for inference at a regular frequency, to provide real-time predictions and hence, support more effective patient monitoring.}

\vspace{2mm}
\noindent
\revise{\powerpoint{\model model.} The core component \model of \framework takes the data as input to generate predictions based on the collaboration of three modules: (1) Time-Invariant Module computes the time-invariant feature importance representation; (ii) Time-Variant Module computes the time-variant feature importance representation with the guidance of Time-Invariant Module; (iii) Prediction Module takes into account the information of both Time-Invariant Module and Time-Variant Module to generate the final predictions. The design details and interaction of these modules will be elaborated in Section~\ref{sec:methodology}.}

\vspace{2mm}
\noindent
\powerpoint{Doctor Validation.} With the \model model, \framework facilitates doctor validation with accurate and interpretable clinical decision support. 
We illustrate three representative scenarios
with the application of hospital-acquired AKI prediction as follows.
\begin{itemize}[leftmargin=*]
    \item \revise{\highlight{Real-time Prediction \& Alert for Daily Consultation.} Suppose Doctor X is monitoring the chance of developing AKI for a hospitalized Patient Y on a daily basis.
    \framework can help Doctor X via feeding the daily generated EMR data of Patient Y for analysis and computing the probability of Patient Y developing AKI in two days. Once the prediction (e.g., $85\%$) exceeds a predefined risk threshold (e.g., $75\%$), \framework will send an alert to notify Doctor X that Patient Y is at risk, such that Doctor X can attend to Patient Y in time to take some preventive treatments to avoid deterioration.
    In this scenario, \framework assists doctors with clinical decision making and contributes to better patient management.}
    
    \item \revise{\highlight{Patient-Level Interpretation for Patient-Level Analysis.}
    Following the example above, suppose Doctor X has suggested some treatments to Patient Y in advance and afterward, Doctor X decides to further investigate the EMR data of Patient Y, to find out why he/she has an $85\%$ probability of developing AKI.
    In such a scenario, \framework can provide interpretation analysis of Patient Y based on the history time-series EMR data, and help identify the specific features of particular visits to the hospital that are responsible for the prediction of developing AKI in the future.
    With such patient-level interpretation, \framework improves the understanding of doctors on each patient, and helps identify the underlying reasons for developing certain diseases such as AKI.}
    
    \item \revise{\highlight{Feature-Level Interpretation for Medical Research.}
    Suppose Doctor X has analyzed the EMR data of a cohort of patients, who have a relatively high probability (e.g., higher than the threshold $75\%$) of developing AKI, and has noticed these patients share certain similarities in the history time-series EMR data based on the patient-level interpretation, e.g., an increasing importance of the laboratory test ``C-Reactive Protein'' in recent examinations. 
    Then Doctor X decides to further investigate the underlying pattern of this laboratory test with regard to the AKI development in the cohort for medical research.
    \framework can support the needs of Doctor X with the feature-level interpretation, which depicts the feature importance changing pattern of this feature over time among all patients. In this way, \framework assists Doctor X in understanding the characteristics of this feature in the AKI development and hence, contributes to the advancement of medical research.}
\end{itemize}

\section{{\Large\model} Model}
\label{sec:methodology}

We denote a time-series sample as $\bm{X}=\{\bm{x_1}, \cdots, \bm{x_T}\}$ of $T$ time windows, where the window length is flexible, e.g., one day, one hour or one visit.
Specifically, for healthcare analytics,
each time window contains the medical features of the patient extracted from his/her EMR data, denoted as $\bm{x_t}\in R^D$, where $D$ is the number of medical features and $t \in \{1, 2, \cdots, T\}$.
Each sample is extracted from a particular patient, while each patient may have more than one sample\footnote{The detailed extraction process will be explained in Section~\ref{subsubsec:datasets and applications}.}. 
Each sample also has a corresponding label $y$ indicating the medical status of the patient.
In this section, we formulate with the binary classification where the label is a binary value\footnote{In the experiments for healthcare analytics of Section~\ref{sec:experiments}, the binary value indicates whether a patient will develop AKI for hospital-acquired AKI prediction or pass away for in-hospital mortality prediction.}.
We note that this formulation can be readily extended to other learning tasks such as regression by replacing the output activation function with a linear function.

\revise{We shall elaborate on the core component of \framework, the \model model for analytics. 
We first introduce the overall architecture of \model and the three modules of \model: Time-Invariant Module, Time-Variant Module and Prediction Module.}
Then we analyze the importance of each feature to the final prediction of \model.

\subsection{{\Large\model} Architecture Overview}
\label{subsec:combination of local and global module}

\begin{figure}
	\centering
	\includegraphics[width=0.6\linewidth]{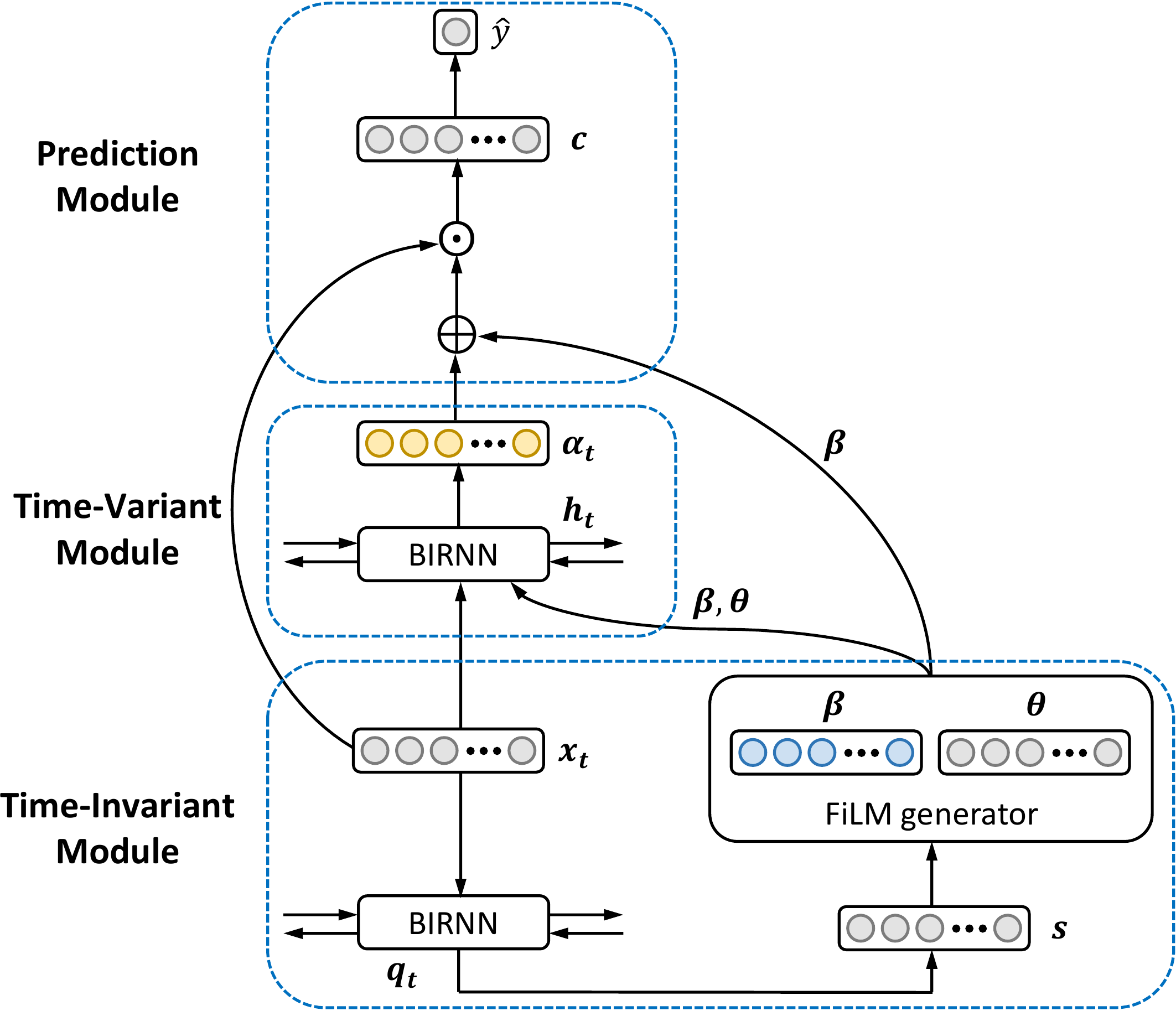}
	\caption{\model with the collaboration of three modules.}
	\label{fig:building block}
\end{figure}

With the input sample $\bm{X}=\{\bm{x_1}, \cdots, \bm{x_T}\}$, we illustrate the generation of the prediction $\hat{y}$ from \model in Figure~\ref{fig:building block}.
Specifically, \model is composed of three modules as follows.

\noindent
\powerpoint{\revise{Time-Invariant Module.}}
For each sample, $\bm{x_t}$ is first fed into a bidirectional RNN (BIRNN)~\cite{schuster1997bidirectional} to compute a hidden representation $\bm{q_t}$.
Then $\bm{q_t}$ from all time windows is averaged into a summary vector $\bm{s}$, which flows to the \film generator, a unit to calculate the scaling parameter $\bm{\beta}$ and the shifting parameter $\bm{\theta}$. 
\revise{This $\bm{\beta}$ calculates the time-invariant feature importance} for the sample whose influence is shared across time windows.

\noindent
\powerpoint{\revise{Time-Variant Module.}}
We design a $BIRNN_{\film}(\cdot)$ with $\bm{\beta}$ and $\bm{\theta}$ \revise{from Time-Invariant Module} to conduct a feature-wise affine transformation over $\bm{x_t}$ of time window $t$.
Then we compute the hidden representation $\bm{h_t}$ with $BIRNN_{\film}(\cdot)$, and feed it to a unit supporting self-attention mechanism to compute \revise{the time-variant feature importance $\bm{\alpha_t}$.}

\noindent
\powerpoint{Prediction Module.}
We aggregate \revise{the time-invariant feature importance $\bm{\beta}$ and the time-variant feature importance $\bm{\alpha_t}$} into an overall feature importance $\bm{\xi_t}$.
Then the final context vector $\bm{c}$ is obtained by summarizing the product of $\bm{\xi_t}$ and $\bm{x_t}$ for each time window $t$. 
Finally, the context vector $\bm{c}$ is used for the prediction of label $\hat{y}$.

\rev{We note that the integration of time-invariant feature importance $\bm{\beta}$ and time-variant feature importance $\bm{\alpha_t}$ is non-trivial, and no previous study has investigated the integration of both.
Specifically, $\bm{\beta}$ computed from Time-Invariant Module will: (i) guide the modulation of the input in Time-Variant Module for calculating $\bm{\alpha_t}$, (ii) integrate with $\bm{\alpha_t}$ in Prediction Module. We shall justify such integration experimentally in Section~\ref{subsubsec:ablation study}.
}

\subsection{{\Large\film}-based Time-Invariant Module}
\label{subsec:film-based global module}

We aim to model the time-invariant feature importance shared across time, where data in all the time windows are required and exploited.
\film is an effective conditioning mechanism that applies a feature-wise affine transformation on the input, and is designed to model the feature importance~\cite{dumoulin2018feature-wise, perez2018film, kim2017dynamic}.
We therefore adopt \film in Time-Invariant Module for computing the time-invariant feature importance, with the input EMR data as the self-conditioning information.

\begin{figure}
\centering
\includegraphics[width=0.8\linewidth]{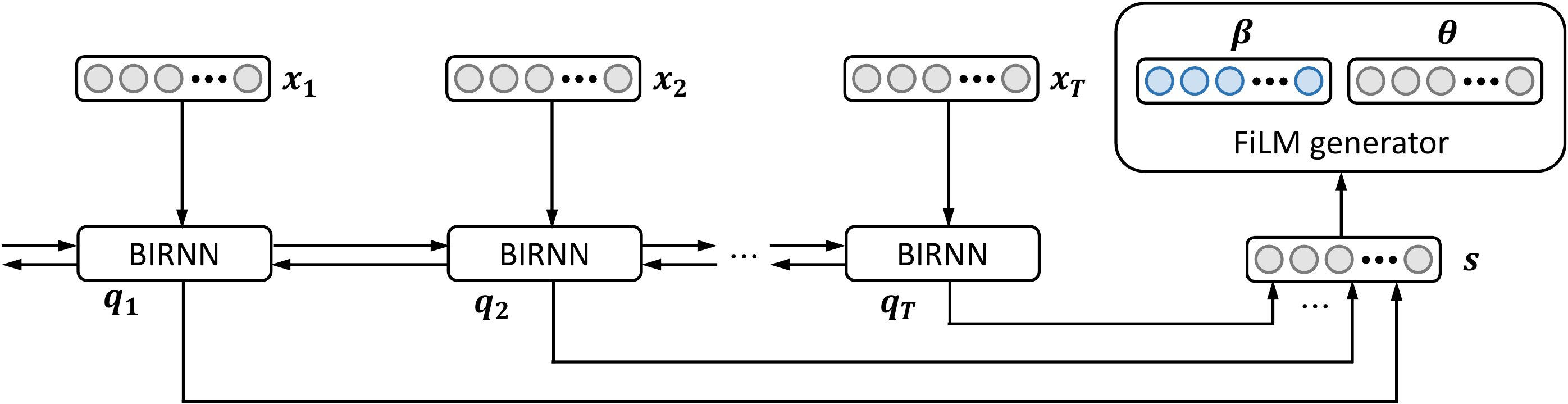}
\caption{Time-Invariant Module of \model.}
\label{fig:global}
\end{figure}

Specifically, with \film, we can obtain the feature-wise scaling parameter and shifting parameter for each sample.
The scaling parameter represents the time-invariant importance of the features across the entire time range of each sample.
The detailed structure of \film-based Time-Invariant Module is illustrated in Figure~\ref{fig:global}.

We first feed the time-series EMR data to a standard BIRNN model and obtain the hidden representations:
\begin{equation}
\label{formula:standard rnn in global}
(\bm{q_1}, \cdots, \bm{q_t}, \cdots, \bm{q_T}) = BIRNN (\bm{x_1}, \cdots, \bm{x_t}, \cdots, \bm{x_T})
\end{equation}
where $BIRNN(\cdot)$ refers to a bidirectional GRU model, 
and the hidden representation $\bm{q_t} = [\overrightarrow{\bm{q_t}}; \overleftarrow{\bm{q_t}}]$ is the concatenation of the hidden states computed from both directions (i.e., forward and backward). Specifically, $\overrightarrow{\bm{q_t}}$ is obtained via a forward GRU model (from $\bm{x_1}$ to $\bm{x_t}$), and $\overleftarrow{\bm{q_t}}$ via a backward GRU model (from $\bm{x_T}$ to $\bm{x_t}$).
The major advantage of BIRNN lies in its capability to capture both the forward and the backward temporal relationship of the EMR data, which is similar to the procedure that doctors analyze the history EMR data of a patient from both directions.
Consequently, BIRNN provides a comprehensive representation of the time-series EMR data.

We further aggregate all the hidden representations of all the time windows into a summary vector $\bm{s}$:
\begin{equation}
\label{formula:summary vector computation}
\bm{s} = \frac{1}{T} \sum_{t=1}^T \bm{q_t}
\end{equation}
Then this aggregated representation $\bm{s}$ flows into \film to calculate the scaling parameter and the shifting parameter:

\begin{equation}
\label{formula:scaling parameter computation}
\bm{\beta} = \bm{W_\beta} \bm{s} + \bm{b_\beta} 
\end{equation}
\begin{equation}
\label{formula:shifting parameter computation}
\bm{\theta} = \bm{W_\theta} \bm{s} + \bm{b_\theta} 
\end{equation}

Note that $\bm{\beta}$ and $\bm{\theta}$ obtained in this module will also serve as auxiliary inputs to \revise{Time-Variant Module} of \model for better predictions, as they guide the modulation of the input EMR data in a feature-wise manner.
Further, the scaling parameter $\bm{\beta}$ determines the scale of each feature and thus indicates the importance of each feature.
For a given sample, $\bm{\beta}$ is shared and fixed through all the time windows.
$\bm{\beta}$ is meant to serve as the \revise{time-invariant feature importance}, which is required and thus integrated into Prediction Module.

\subsection{\revise{Attention-based Time-Variant Module}}
\label{subsec:attention-based local module}
\begin{figure}
\centering
\includegraphics[width=0.7\linewidth]{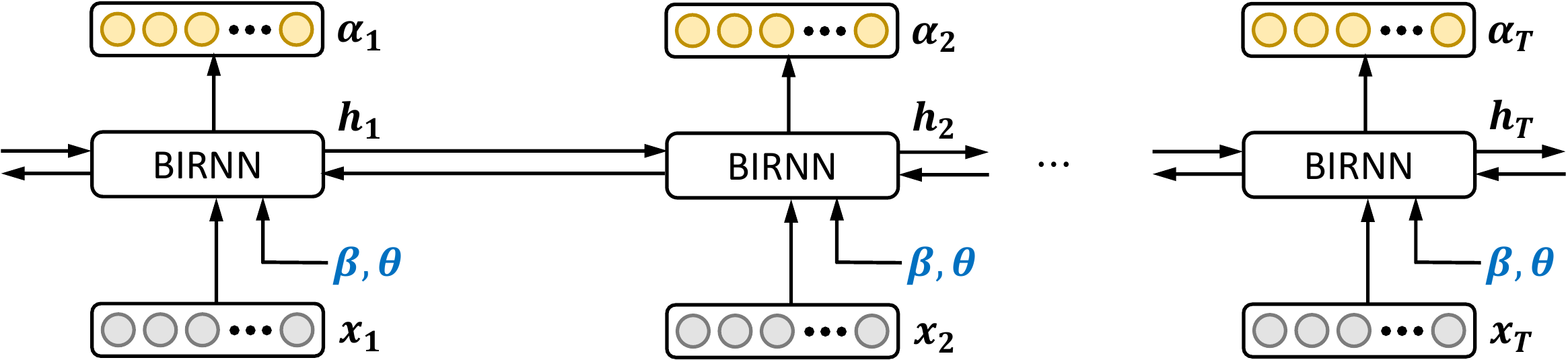}
\caption{Time-Variant Module of \model.}
\label{fig:local}
\end{figure}

We aim to differentiate the influence of different features \revise{in different time windows when modeling the time-variant feature importance.}
Similar tasks have been successfully supported with the adoption of the self-attention mechanism in many areas~\cite{cheng2016long, xu2015show}.
We therefore introduce the self-attention mechanism to Time-Variant Module (as illustrated in Figure~\ref{fig:local}) to calculate the time-variant feature importance specific to each time window.

We first feed the time-series data into an adapted BIRNN model with the
auxiliary information from Time-Invariant Module to calculate time-variant hidden representations:

\begin{equation}
\label{formula:film rnn in local}
(\bm{h_1}, \cdots, \bm{h_t}, \cdots, \bm{h_T}) = BIRNN_{\film} (\bm{x_1}, \cdots, \bm{x_t}, \cdots, \bm{x_T}; \bm{\beta}, \bm{\theta})
\end{equation}
where $BIRNN_{\film}(\cdot)$ refers to the \film-based bidirectional GRU model, which also takes the scaling parameter $\bm{\beta}$ and the shifting parameter $\bm{\theta}$ from \revise{Time-Invariant Module} as additional input.

Specifically, the detailed transformation of $BIRNN_{\film}(\cdot)$ is given as follows, where the revised update gate $\bm{z_t}$, reset gate $\bm{r_t}$, temporary hidden state $\bm{\widetilde{h}_t}$ and final hidden state $\bm{h_t}$ are calculated in a bidirectional GRU model in order:

\begin{equation}
\label{formula:revised update gate}
\bm{z_t} = \sigma (\film (\bm{W_z} \bm{x_t}; \bm{\beta}, \bm{\theta}) + \bm{U_z} \bm{h_{t-1}})
\end{equation}
\begin{equation}
\label{formula:revised reset gate}
\bm{r_t} = \sigma (\film (\bm{W_r} \bm{x_t}; \bm{\beta}, \bm{\theta}) + \bm{U_r} \bm{h_{t-1}})
\end{equation}
\begin{equation}
\label{formula:revised temporary hidden state}
\bm{\widetilde{h}_t} = \tanh (\film (\bm{\widetilde{W}} \bm{x_t}; \bm{\beta}, \bm{\theta}) + \bm{r_t} \odot \bm{\widetilde{U}} \bm{h_{t-1}})
\end{equation}
\begin{equation}
\label{formula:revised hidden state}
\bm{h_t} = (\bm{1} - \bm{z_t}) \odot \bm{\widetilde{h}_t} + \bm{z_t} \odot \bm{h_{t-1}}
\end{equation}
where $\sigma(\cdot)$ is the $sigmoid(\cdot)$ activation function and ``$\odot$'' denotes the element-wise multiplication.
Different from the standard bidirectional GRU model, $BIRNN_{\film}(\cdot)$ also exploits $\bm{\beta}$ and $\bm{\theta}$ from \revise{Time-Invariant Module} with a feature-wise affine transformation $\film(\cdot)$ defined as:
\begin{equation}
\label{formula:feature-wise transformation definition}
\film(\bm{x};\bm{\beta}, \bm{\theta}) = \bm{\beta} \odot \bm{x} + \bm{\theta}
\end{equation}

We then employ a self-attention mechanism to compute the \revise{time-variant feature importance}:

\begin{equation}
\label{formula:local attention}
\bm{\alpha_t} = \tanh (\bm{W_\alpha} \bm{h_t} + \bm{b_\alpha})
\end{equation}
where $\bm{\alpha_t}$ of the time window $t$ will be fed into Prediction Module to attend to the features of $\bm{x_t}$ for prediction.

\subsection{Prediction Module}
\label{subsec:prediction}
\begin{figure}
\centering
\includegraphics[width=0.7\linewidth]{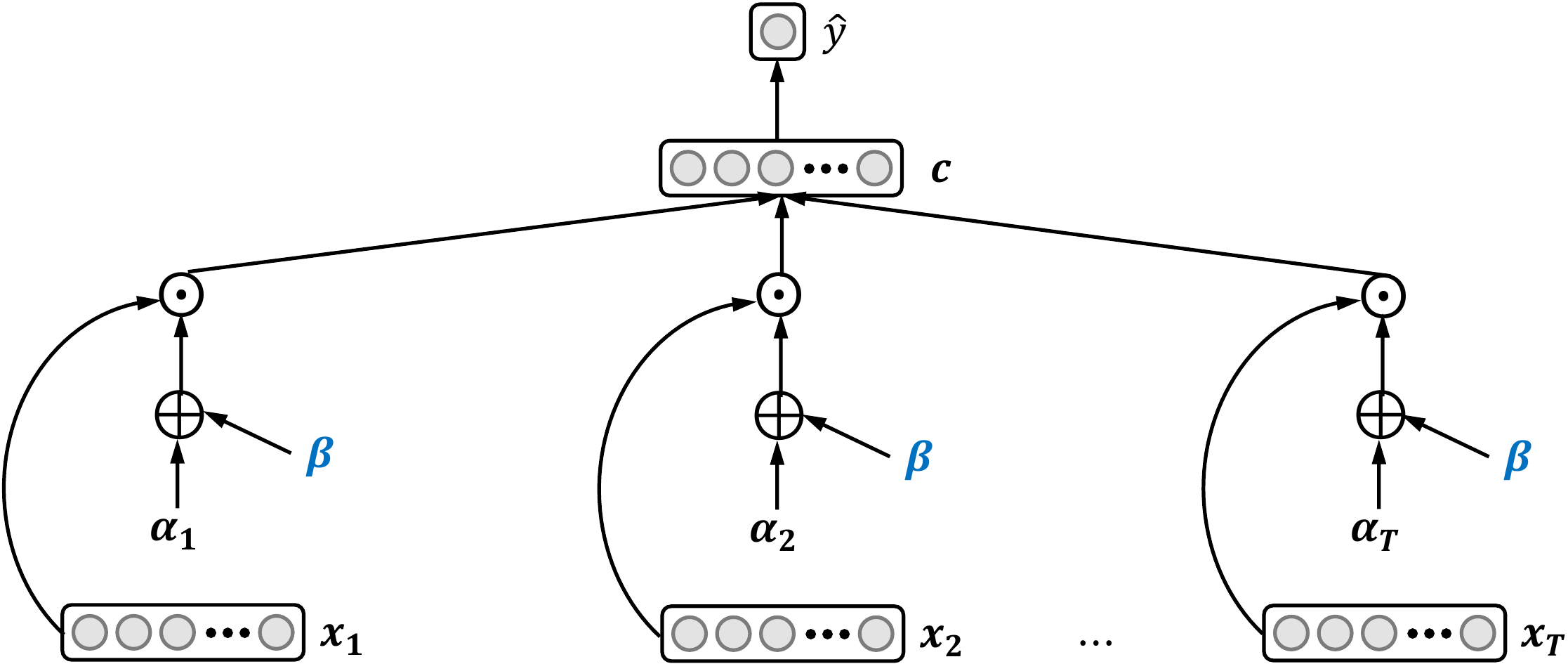}
\caption{Prediction Module of \model.}
\label{fig:prediction module}
\end{figure}

In Prediction Module, we produce the final prediction of \model illustrated in Figure~\ref{fig:prediction module}.
After obtaining the scaling parameter $\bm{\beta}$ \revise{as the time-invariant feature importance and the self-attention $\bm{\alpha_t}$ as the time-variant feature importance}, we obtain the overall influence in Prediction Module:

\begin{equation}
\label{formula:combine local and global importance}
\bm{\xi_t} = \bm{\beta} \oplus \bm{\alpha_t}
\end{equation}
where ``$\oplus$'' denotes the element-wise summation. \rev{Note that $\bm{\beta}$ and $\bm{\alpha_t}$ are intermediate neural network outputs which should be combined by direct summation for generally better results~\cite{wang2016deeply, he2016identity, parikh2016decomposable}.
}
Thereby, \revise{we integrate both the general time-invariant feature importance $\bm{\beta}$ and the fine-grained time-variant feature importance $\bm{\alpha_t}$} together into the final feature importance representation $\bm{\xi_t}$. 

We then obtain the context vector $\bm{c}$ by aggregating the element-wise product of $\bm{\xi_t}$ and the corresponding input $\bm{x_t}$ at each time window $t$:
\begin{equation}
\label{formula:context vector}
\bm{c} = \sum_{t=1}^T \bm{\xi_t} \odot \bm{x_t}
\end{equation}

The final predicted label of \model is therefore:

\begin{equation}
\label{formula:final prediction}
\hat{y} = \sigma (\langle \bm{w}, \bm{c} \rangle + b)
\end{equation}
where ``$\langle\cdot, \cdot\rangle$'' denotes the inner-product of two vectors.
Finally, the training of the whole framework is achieved via the optimization of a predefined loss function $\mathcal{L}(\hat{y}, y)$ between the prediction $\hat{y}$ and the ground truth label $y$, e.g., the cross-entropy loss function for binary classification:

\begin{equation}
\label{formula:loss function}
\mathcal{L}(\hat{y}, y) = -y \cdot \log (\hat{y}) - (1-y) \cdot \log (1 - \hat{y})
\end{equation}
Specifically, stochastic gradient descent back-propagation optimization can be employed to train \model's model parameters $\Theta=\{\bm{W_\beta}, \bm{b_\beta}, \bm{W_\theta}, \bm{b_\theta}, \bm{W_z}, \bm{U_z}, \bm{W_r}, \bm{U_r}, \bm{\widetilde{W}}, \bm{\widetilde{U}}, \bm{W_\alpha}, \bm{b_\alpha}, \bm{w}, b\}$ in an end-to-end manner.

\subsection{Feature Importance for Interpretation}
\label{subsec:contribution of a feature to y}

In this subsection, we analyze the feature importance of $x_{t,d}$ to the predicted label $\hat{y}$, where $x_{t,d}$ denotes the value of feature $d$ at time window $t$ of a sample $\bm{X}$.

In binary classification, $\hat{y}$ corresponds to the probability of the sample to be classified as the positive class.
Expanding Equation~\ref{formula:final prediction} with $\bm{c}$ (Equation~\ref{formula:context vector}) and $\bm{\xi_t}$ (Equation~\ref{formula:combine local and global importance}), we have:

\begin{equation}
\label{formula:contribution analysis}
\begin{split}
& \hat{y} = \sigma (\langle \bm{w}, \bm{c} \rangle + b) \\
& = \sigma (\langle \bm{w}, \sum_{t=1}^T \bm{\xi_t} \odot \bm{x_t} \rangle + b) \\
& = \sigma (\langle \bm{w}, \sum_{t=1}^T {(\bm{\beta} \oplus \bm{\alpha_t})} \odot \bm{x_t} \rangle + b) \\
& = \sigma (\sum_{t=1}^T \langle \bm{w},  {(\bm{\beta} \oplus \bm{\alpha_t})} \odot \bm{x_t} \rangle + b) \\
\end{split}
\end{equation}

Then the feature importance of $x_{t,d}$ to the prediction $\hat{y}$ can be derived as follows:

\begin{equation}
\label{formula:contribution of a feature}
FI(\hat{y}, x_{t,d}) = (\beta_d + \alpha_{t,d}) \cdot w_d
\end{equation}
where $\beta_d$, $\alpha_{t,d}$ and $w_d$ correspond to the $d$-th element of $\bm{\beta}$, $\bm{\alpha_t}$ and $\bm{w}$ respectively.
We can observe that both $\beta_d$ and $\alpha_{t,d}$ directly influence the feature importance of $x_{t,d}$ to $\hat{y}$.

Substitute Equation~\ref{formula:contribution of a feature} into Equation~\ref{formula:contribution analysis}, we have:
\begin{equation}
\label{formula:contribution of all features}
\hat{y} = \sigma (\sum_{t=1}^T \sum_{d=1}^D FI(\hat{y}, x_{t,d}) \cdot x_{t,d} + b)
\end{equation}
which demonstrates that all the features contribute to the final prediction of \model with the corresponding feature importance given by $FI(\hat{y}, x_{t,d})$.
Further, an $x_{t,d}$ with a positive feature importance value indicates that $x_{t,d}$ positively contributes to the final prediction, while a negative feature importance value the opposite.
\rev{
We note that \framework also takes into account the feature interactions during this process, specifically interactions of the input features are first modeled using BIRNN in both Time-Invariant Module and Time-Variant Module, then captured in $\bm{\beta}$ and $\bm{\alpha_t}$, and finally, integrated in $FI(\hat{y}, x_{t,d})$.
}

\rev{With $FI(\hat{y}, x_{t,d})$, \framework can provide interpretable decision support for high stakes applications.
Specifically, in healthcare analytics, \framework can help reveal the $FI(\hat{y}, x_{t,d})$ changing pattern of features over time for each patient to show the influence of the varying time-series features, i.e., support patient-level interpretation analysis.
Therefore, \framework can assist doctors to pinpoint the underlying problems of the patient.
For each feature, \framework can help identify the $FI(\hat{y}, x_{t,d})$ changing pattern over time on a cohort of patients, and further facilitate the understanding of the development of certain diseases for doctors/clinicians, i.e., support feature-level interpretation analysis.}

\section{Experiments}
\label{sec:experiments}

\subsection{Experimental Set-Up}
\label{subsec:experimental set-up}
\subsubsection{Datasets and Applications}
\label{subsubsec:datasets and applications}

We evaluate \revise{\framework} in two real-world longitudinal EMR datasets, the NUH-AKI dataset and the MIMIC-III dataset.

\begin{figure}
	\centering
	\includegraphics[width=0.7\linewidth]{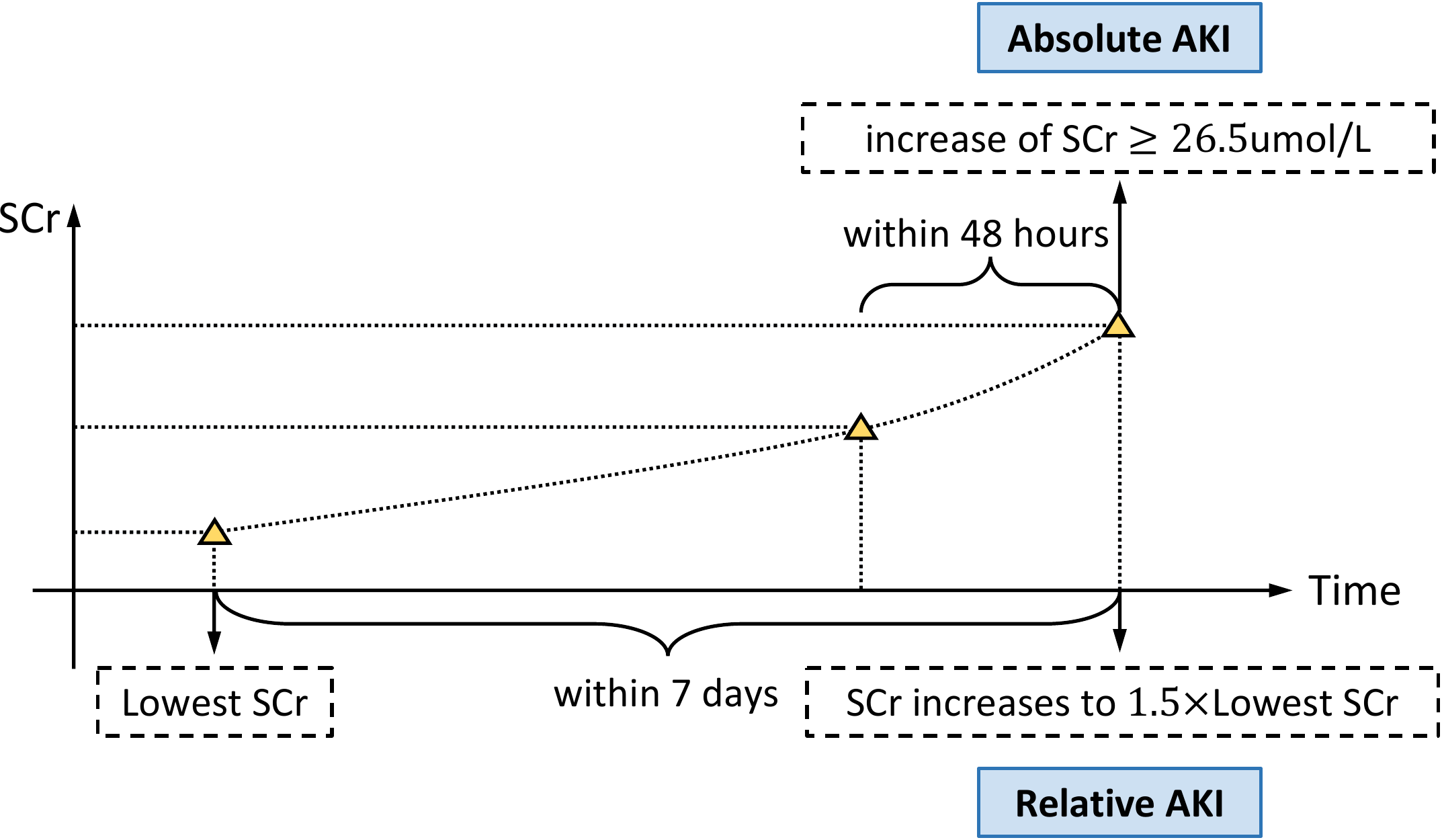}
	\caption{AKI detection criteria: (i) absolute AKI, and (ii) relative AKI.
	}
	\label{fig:aki detection}
\end{figure}

\vspace{1mm}
\noindent
\powerpoint{NUH-AKI Dataset} a sub-dataset extracted from all EMR data in National University Hospital in Singapore recording more than 100,000 patients' \revise{EMR data} in 2012.
\revise{In this dataset,} we target at \powerpoint{hospital-acquired AKI prediction}, i.e., to predict if a patient will develop AKI in a hospitalized admission.
As explained by medical experts, AKI disease is defined according to the KDIGO criterion~\cite{kellum2012kidney}.
The definition of AKI is based on a laboratory test serum creatinine (SCr) and there are two AKI detection criteria, absolute AKI and relative AKI (as illustrated in Figure~\ref{fig:aki detection}).
Absolute AKI refers to the situation when the SCr value increases by over $26.5 umol/L$ within the past $48$ hours, while relative AKI refers to the case when the SCr value increases to more than $1.5$ times of the lowest SCr value within the past seven days.
For each hospitalized admission of a patient, both AKI detection criteria are checked in order to derive the label of this admission, and either criterion can cause the AKI label to be positive.

\begin{figure}
	\centering
	\includegraphics[width=0.7\linewidth]{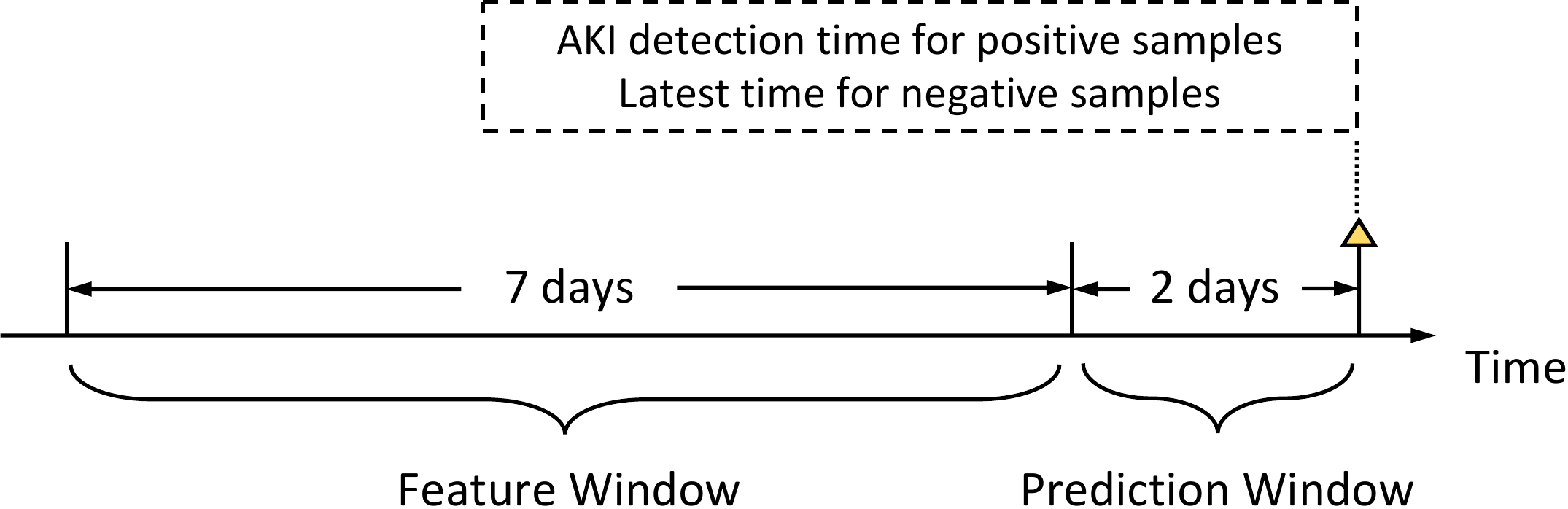}
	\caption{Feature Window and Prediction Window in hospital-acquired AKI prediction.
	}
	\label{fig:feature window and prediction window}
\end{figure}

In the NUH-AKI dataset, each hospitalized admission is used as a sample.
If a sample is positive, i.e., the patient develops AKI in this admission, we record the time when AKI is detected, trace two days back in time as ``Prediction Window'' which is not used and continue to trace seven more days back in time as ``Feature Window'' which is used to construct $\bm{x}$.
Otherwise, if a sample is negative without developing AKI, then the time recorded for the latest medical feature in this admission is used to determine the Prediction Window and Feature Window. 
The relationship between Feature Window and Prediction Window is shown in Figure~\ref{fig:feature window and prediction window}.
In hospital-acquired AKI prediction, we utilize the time-series laboratory tests of each sample in Feature Window as input to predict if the patient will develop AKI in this admission in two days.

\vspace{1mm}
\noindent
\powerpoint{MIMIC-III Dataset}~\cite{johnson2016mimic} is a public dataset spanning from 2001 to 2012, recording EMR data for more than $40,000$ ICU patients admitted to the critical care units.
In this MIMIC-III dataset, we conduct \powerpoint{in-hospital mortality prediction} with the time-series laboratory tests as input.
Specifically, each admission corresponds to one visit of each patient to the hospital and for each admission, if the patient stays in the hospital for more than 48 hours, we use it as one sample. The mortality label is derived by checking whether the patient passes away in the hospital in this admission or not. \revise{Then the corresponding} time-series $\bm{x}$ is extracted from the laboratory tests of this admission.

We summarize some important statistics for both datasets in Table~\ref{tab:dataset statistics}.
\rev{
Specifically, we divide Feature Window into a number of time windows by the window length and average the value of the same laboratory test for each time window, which is a typical way to transform EMR data for analytics~\cite{zhou2014micro, che2015deep, lipton2015learning}.
}
\rev{
Then we conduct feature normalization on the laboratory test value $x$ to obtain the normalized value $x'$ as the input for analytics, specifically $x' = (x - min)/(max -min)$.
We note that while the laboratory tests used in the experiments are numerical features, \framework can readily deal with categorical or discrete features by transforming them into numerical features via standard preprocessing steps (e.g., sklearn.\-preprocessing.\-OneHotEncoder~\cite{sklearn}, pandas.\-get\_dummies~\cite{pandas}) before feeding them as input.
}

\begin{table}[t]
	\small
	\centering
	\caption{Dataset Statistics}
	\label{tab:dataset statistics}
	\begin{tabular}{c|c|c}
		\toprule[2pt]
		Dataset & NUH-AKI & MIMIC-III \\ \midrule[1pt]
		Feature Number & $709$ & $428$ \\ \midrule[0.1pt]
		Sample Number & $20732$ & $51826$ \\ \midrule[0.1pt]
		Positive Sample Number & $911$ & $4280$ \\ \midrule[0.1pt]
		Negative Sample Number & $19821$ & $47546$ \\ \midrule[0.1pt]
		Feature Window Length & $7$ days & $48$ hours \\ \midrule[0.1pt]
		Time Window Length & $1$ day & 2 hours \\ \midrule[0.1pt]
		Time Window Number & $7$ & $24$ \\
		\bottomrule[2pt]
	\end{tabular}
\end{table} 

\subsubsection{Baseline Methods}
\label{subsubsec:baseline methods}
We compare \revise{\framework with LR, \rev{Gradient Boosting Decision Tree (GBDT)}, the standard BIRNN and several state-of-the-art methods including RETAIN~\cite{choi2016retain} and variants of Dipole~\cite{ma2017dipole}.} The details of these baselines are as follows.

\begin{itemize}[leftmargin=*]
\item \powerpoint{LR} takes the aggregated time-series EMR data as input for prediction. The aggregation operation calculates the average value of the same feature across the time series.

\rev{\item \powerpoint{GBDT} is an ensemble model composed of decision trees, which also takes the aggregated time-series EMR data as input.}

\item \powerpoint{BIRNN} takes time-series EMR data as input and uses the BIRNN's last hidden state for prediction.

\item \powerpoint{RETAIN}~\cite{choi2016retain} is a reverse time attention model
which devises a two-level attention mechanism, i.e., visit-level attention and feature-level attention to facilitate interpretability. 

\item \powerpoint{Dipole}~\cite{ma2017dipole} is an attention-based BIRNN model
which can achieve the interpretability for each visit through three different attention mechanisms as follows.

\item \powerpoint{Dipole$_{loc}$} is Dipole with a location-based attention mechanism in which the attention weights are computed solely based on the current hidden state.

\item \powerpoint{Dipole$_{gen}$} is Dipole with a general attention mechanism in which a matrix is used to capture the relationship between every two hidden states.

\item \powerpoint{Dipole$_{con}$} is Dipole with a concatenation-based attention mechanism in which the attention weights are computed from the concatenation of the current hidden state and each previous hidden state.
\end{itemize}

For the experiments, we randomly partition the samples into $80\%$, $10\%$ and $10\%$ for training, validation and test respectively. During training, for each approach (either \framework or baselines), the hyperparameters which can achieve the best performance in the validation data are chosen and then applied to the test data for reporting experimental results.
For both applications formalized as binary classification, we choose the area under the ROC curve (AUC), as well as the mean cross-entropy loss (CEL) per sample as evaluation metrics, and an accurate prediction model should have a high AUC value but a low CEL value. Then we report the AUC value and the CEL value averaged of $10$ repeats in the test data.

\begin{figure}
	\centering
	\includegraphics[width=0.8\linewidth]{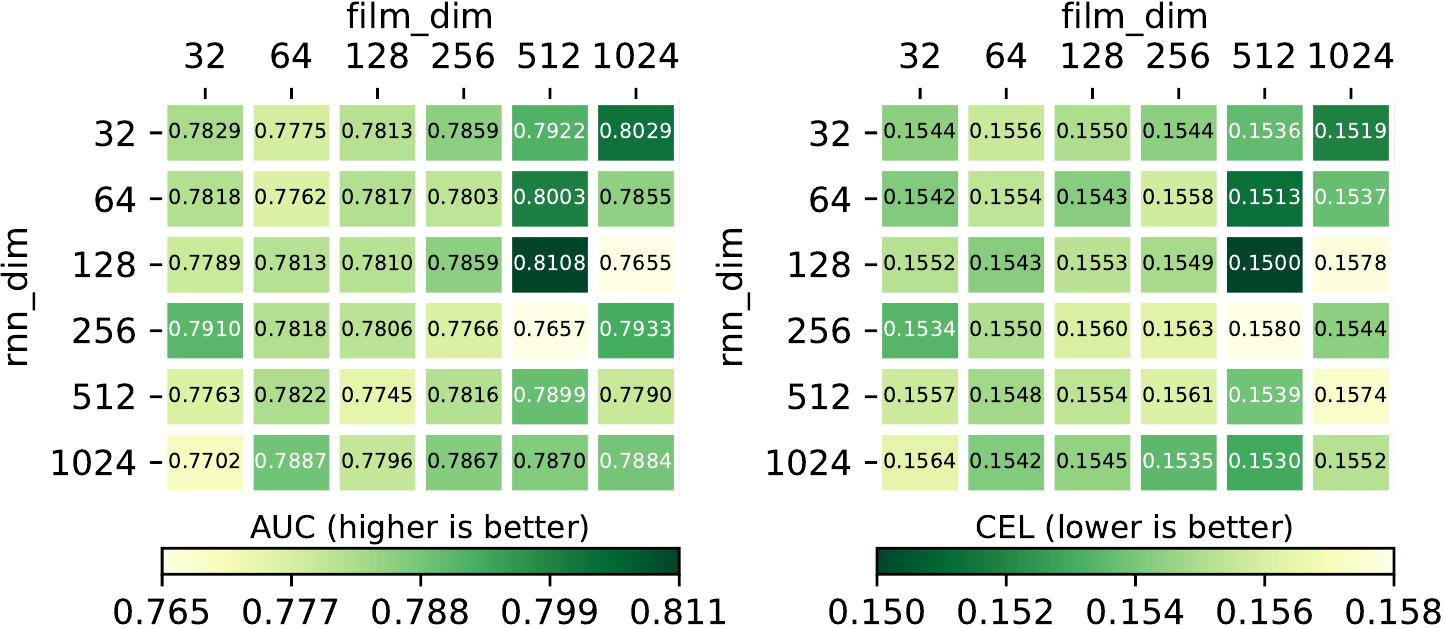}
	\caption{\revise{Sensitivity analysis of \framework on \highlight{rnn\_dim} and \highlight{film\_dim} in the NUH-AKI dataset}.}
	\label{fig:aki hyperparameter sensitivity analysis}
\end{figure}

\begin{figure}
	\centering
	\includegraphics[width=0.8\linewidth]{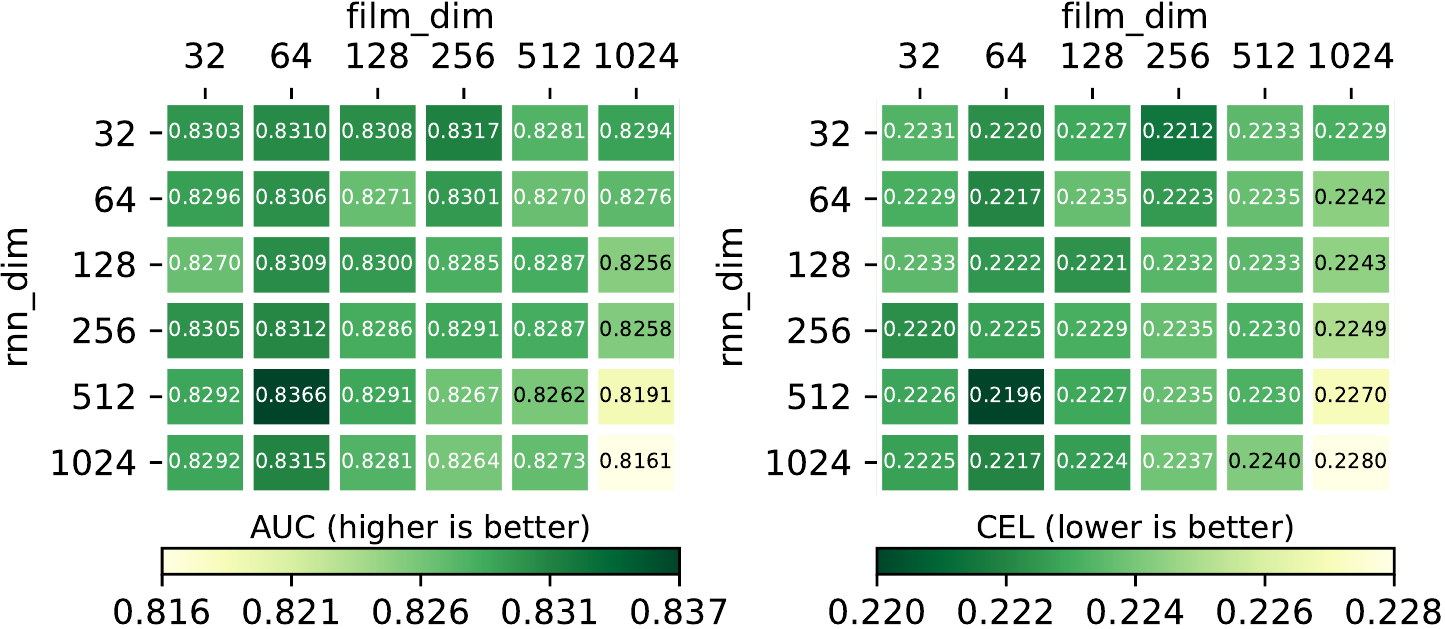}
	\caption{\revise{Sensitivity analysis of \framework on \highlight{rnn\_dim} and \highlight{film\_dim} in the MIMIC-III dataset}.}
	\label{fig:mimic hyperparameter sensitivity analysis}
\end{figure}

\rev{Specifically, for \framework, we conduct the sensitivity analysis on two critical hyperparameters: (i) \highlight{rnn\_dim}, the dimension of $\bm{h_t}$ in Time-Variant module; and (ii) \highlight{film\_dim}, the dimension of $\bm{q_t}$ in Time-Invariant Module. Both hyperparameters are tuned in the range $\{32, 64, 128, 256, 512, 1024\}$.
Figure~\ref{fig:aki hyperparameter sensitivity analysis} and Figure~\ref{fig:mimic hyperparameter sensitivity analysis} illustrate the results of different \highlight{rnn\_dim} and \highlight{film\_dim} settings in both datasets.}
Based on the results, we adopt the best-performing hyperparameter setting \highlight{rnn\_dim}=128, \highlight{film\_dim}=512 in the NUH-AKI dataset and \highlight{rnn\_dim}=512, \highlight{film\_dim}=64 in the MIMIC-III dataset.
\rev{Other hyperparameters include the learning rate $0.001$, the weight decay $5\times10^{-5}$, and the epoch number $200$ with early stopping.}

\subsubsection{\rev{Experimental Environment}}
\label{subsubsec:experimental environment}
\rev{The experiments are conducted in a server with 2 x Intel Xeon Silver 4114 (2.2GHz, 10 cores), 256G memory, 8 x GeForce RTX 2080 Ti. Models are implemented in PyTorch 1.3.1 with CUDA 10.2.}

\subsection{Prediction Results}
\label{subsec:prediction results}
\subsubsection{Comparison with Baseline Methods}
\label{subsubsec:comparison with baseline methods}
\begin{figure}
\centering
\includegraphics[width=0.8\linewidth]{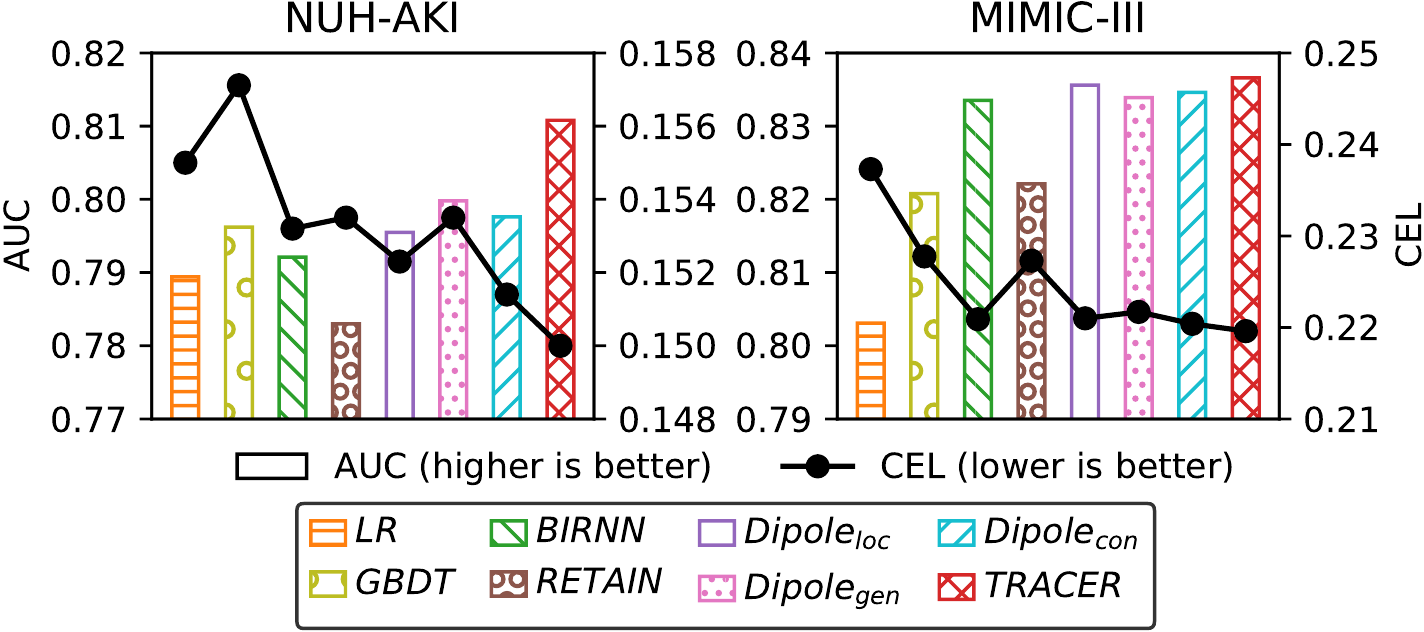}
\caption{\rev{Experimental results of LR, GBDT, BIRNN, RETAIN, Dipole$_{loc}$, Dipole$_{gen}$, Dipole$_{con}$ and \framework.}}
\label{fig:comparison with baseline methods}
\end{figure}

\rev{We report the experimental results of LR, GBDT, BIRNN, RETAIN, Dipole with three different attention mechanisms and our \framework in Figure~\ref{fig:comparison with baseline methods}.}
In both applications, \revise{\framework} achieves the highest AUC values \revise{and the lowest CEL values}, confirming that \revise{the time-invariant and the time-variant feature importance jointly} contribute to the modeling of the time-series EMR data and hence, both are essential to improve the prediction performance.

\rev{In both datasets, \framework outperforms LR and GBDT significantly and the superiority of \framework results from the capability of utilizing time-series data for analytics.}

Compared with RETAIN, \revise{\framework} can achieve a higher AUC value \revise{and a lower CEL value} with a large margin. The reason may be two-fold. First, RETAIN incorporates time-series EMR data in the reverse time order and therefore, loses the forward time-series information in the data. Second, \revise{in the \model model of \framework}, the devised \film mechanism which can \revise{capture the time-invariant feature importance as general guidance to the model learning}, poses a positive influence to improving the performance of \framework.  

Compared with BIRNN, \revise{\framework} illustrates better prediction performance in both datasets, in terms of AUC and CEL. As for interpretability, \revise{\framework} can explain the prediction results via \revise{the feature importance}, whereas BIRNN is hard to interpret.

Compared with Dipole, \revise{\framework} exhibits an obvious advantage over Dipole in the NUH-AKI dataset, and also outperforms Dipole in the MIMIC-III dataset.
From the perspective of interpretability, \revise{\framework} models the feature importance, which is more informative than the visit importance modeled by Dipole. 

\rev{In summary, the proposed \framework outperforms LR and GBDT significantly due to the capability of modeling the time-series data} and outperforms RETAIN \revise{by capturing both time-invariant and time-variant feature importance}. Further, \revise{\framework} achieves better interpretability and meanwhile prediction performance than BIRNN and Dipole in both datasets.

\begin{figure}
\centering
\includegraphics[width=0.8\linewidth]{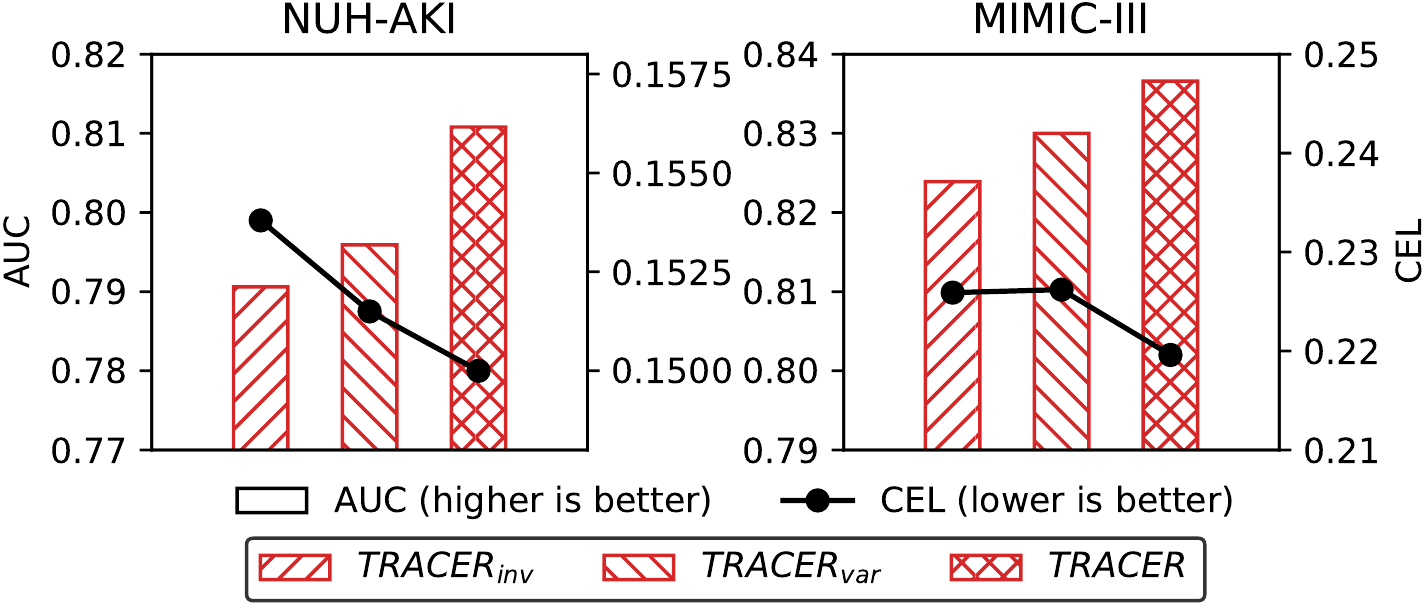}
\caption{\revise{Experimental results of the ablation study}.}
\label{fig:ablation study results}
\end{figure}

\subsubsection{Ablation Study}
\label{subsubsec:ablation study}

We conduct an ablation study \revise{of the \model model in \framework} and illustrate the experimental results in Figure~\ref{fig:ablation study results}.
\revise{$\framework_{inv}$ keeps Time-Invariant Module and Prediction Module of \model for analytics while removing Time-Variant Module to evaluate the influence of Time-Invariant Module.
Similarly, $\framework_{var}$ uses Time-Variant Module and Prediction Module of \model for analytics in order to demonstrate the influence of Time-Variant Module.}

According to Figure~\ref{fig:ablation study results}, we observe that \revise{Time-Invariant Module and Time-Variant Module both contribute to boosting the performance of \framework, as $\framework_{inv}$ and $\framework_{var}$ both achieve a lower AUC \revise{and a higher CEL} than \framework.}

However, compared with Time-Invariant Module, Time-Variant Module poses a larger influence to the performance of \framework, as $\framework_{var}$ achieves a higher AUC value than $\framework_{inv}$ in both datasets.
When considering CEL, $\framework_{var}$ outperforms $\framework_{inv}$ in the NUH-AKI dataset, and performs similarly compared with $\framework_{inv}$ in the MIMIC-III dataset.
This indicates the vital influence of the time-variant feature importance in these two applications.

\subsubsection{\rev{Scalability Test}}
\label{subsubsec:scalability test}

\begin{figure}
	\centering
	\includegraphics[width=0.8\linewidth]{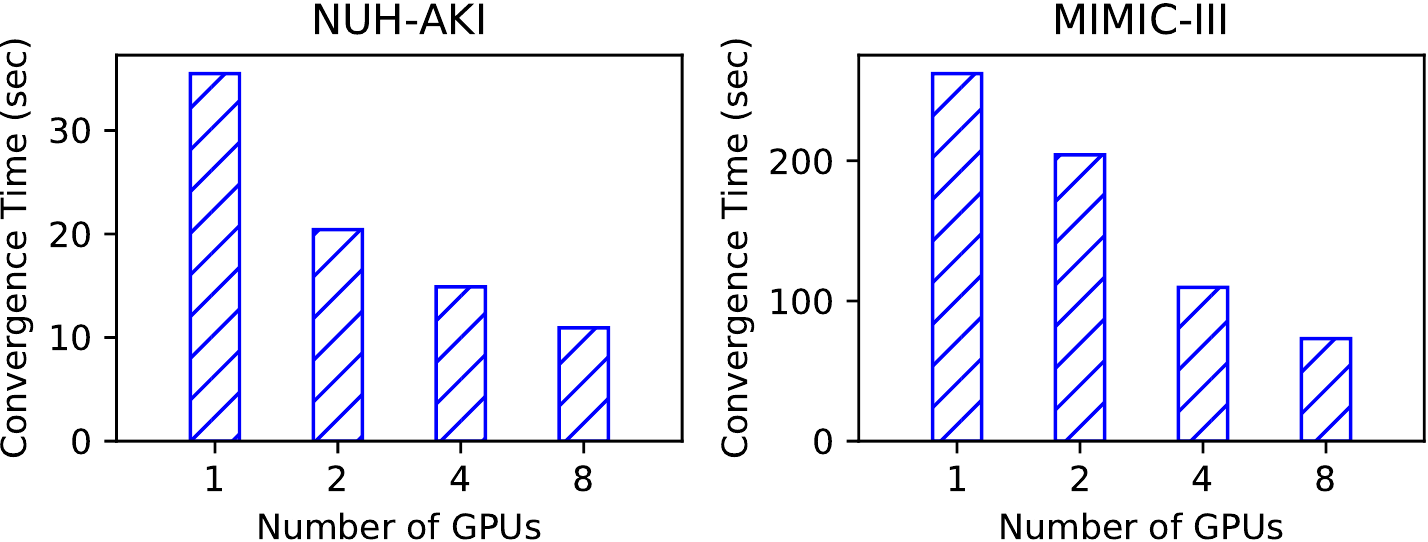}
	\caption{\rev{Experimental results of the scalability test.}}
	\label{fig:scalability test results}
\end{figure}

\rev{
	We study the scalability of \framework by measuring the model convergence time under different number of GPUs in both the NUH-AKI dataset and the MIMIC-III dataset. As can be observed in Figure~\ref{fig:scalability test results}, the convergence time of the NUH-AKI dataset decreases sub-linearly with respect to the number of GPUs used for training, due to the fact that the controlling\footnote{It includes the gradient aggregation among GPUs, the best checkpoint selection and saving, etc,
	which cannot be accelerated with more GPUs.}
	cost becomes more obvious when the processing in each GPU decreases substantially.
    In contrast, when the controlling cost becomes less dominant, we can observe that \framework  yields higher training efficiency as a whole and achieves better scalability for the larger MIMIC-III dataset.
	The study confirms the scalability of \framework that it
	can be accelerated and scaled appropriately with more GPUs depending on the data size and training requirements.
	}

\subsection{\revise{Patient-Level Interpretation}}
\label{subsec:patient-level interpretation results}

In this subsection, we report the patient-level interpretation results to demonstrate how \framework assists doctors to understand why a certain patient develops AKI in the hospital-acquired AKI prediction or passes away in the in-hospital mortality prediction in the patient-level analysis.
\rev{We first adopt the best-performing checkpoint of the \model model in \framework. Then for each patient, we visualize the patient-level interpretation results by plotting the Feature Importance - Time Window distribution
of the features which are identified by doctors to be informative during the diagnosis process.}
Time Window is within Feature Window, ranging from $1$ day to $7$ days in the NUH-AKI dataset and from $2$ hours to $48$ hours in the MIMIC-III dataset respectively.

\subsubsection{\revise{AKI Prediction in the NUH-AKI Dataset}}
\label{subsubsec:aki prediction in hospitalx-aki dataset-patient level}

\begin{figure}
\centering
\includegraphics[width=0.8\linewidth]{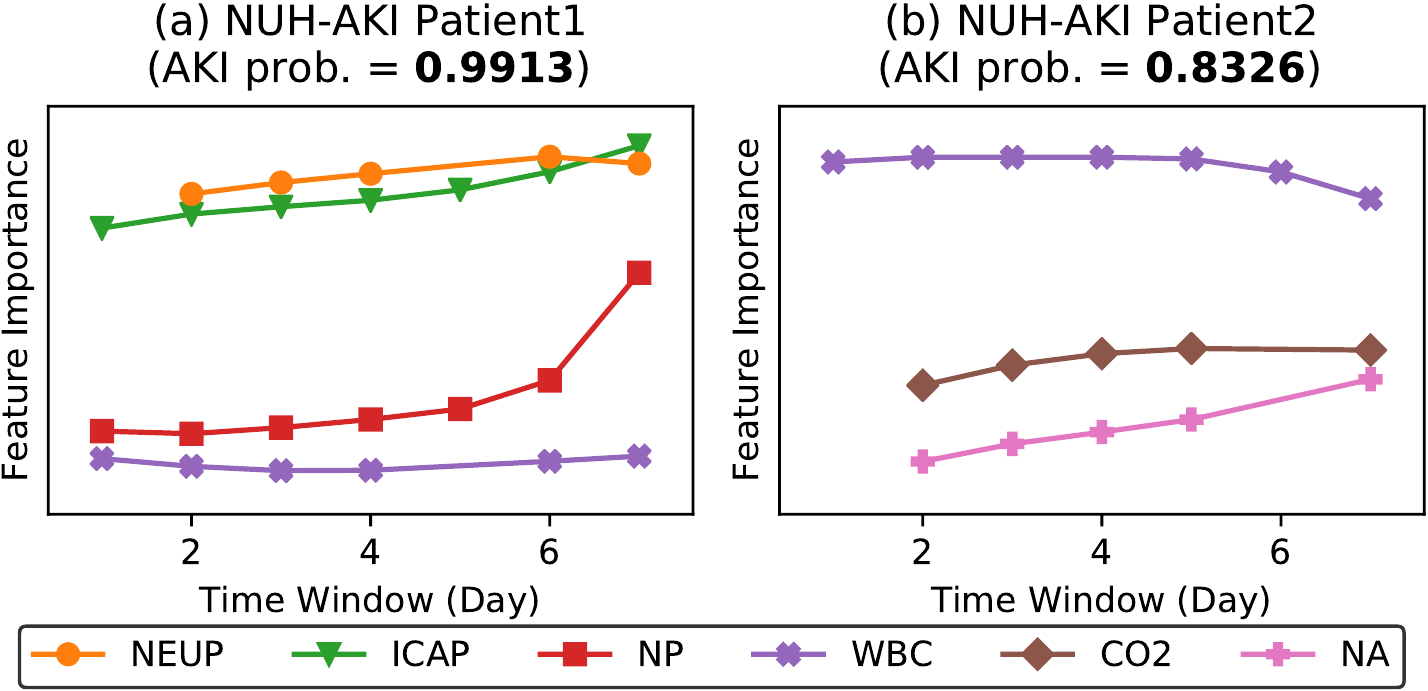}
\caption{\revise{Patient-level interpretation results of \framework 
in the NUH-AKI dataset. 
}}
\label{fig:aki patient-level interpretation results}
\end{figure}

\revise{In the hospital-acquired AKI prediction, we show the patient-level interpretation results of \framework for two representative patients who developed AKI after 48 hours in Figure~\ref{fig:aki patient-level interpretation results}, with the features involved: ``Neutrophils \%'' (\NEUP), ``Ionised CA, POCT'' (\ICAP), ``Sodium, POCT'' (\NP), ``White Blood Cell'' (\WBC), ``Carbon Dioxide'' (\CarbonDioxide) and ``Serum Sodium'' (\NA).}

\revise{As shown in Figure~\ref{fig:aki patient-level interpretation results} (a), in Patient1's interpretation results provided by \framework, we observe that \NEUP feature shows increasingly higher importance along with time, and \WBC is of stable importance. These suggest that Patient1 is suffering from worsening inflammation or infection which both \NEUP and \WBC respond to, although they exhibit different Feature Importance changing patterns.
Then we find that \ICAP and \NP, two kinds of ionised electrolytes in the human body, have a Feature Importance increasing with time. Due to their medical functionality, such as hypocalcemia in AKI and adverse outcomes~\cite{afshinnia2013effect}, and dysnatremia with kidney dysfunction~\cite{gao2019admission}, we presume that Patient1 is developing a worsening electrolyte imbalance along with worsening infection, and thus at high risk of progressing to AKI soon.}

\revise{Then for Patient2 illustrated in Figure~\ref{fig:aki patient-level interpretation results} (b), a relatively high and stable Feature Importance of \WBC is observed, which indicates the presence of inflammation or infection.
Besides, \CarbonDioxide's (bicarbonate) Feature Importance is also on the increase, which is explained by acidosis that builds up with progressive kidney dysfunction~\cite{black1969the}, or worsening lactic acidosis with circulatory shock and end-organ injury including AKI~\cite{mizock1992lactic}.
Once again, the rising Feature Importance changing pattern of \NA represents progressive \NA-fluid imbalance and worsening kidney function in Patient2~\cite{black1969the, gao2019admission}.}

These findings suggest that the patient-level interpretation results of \framework are valuable and vital for doctors to identify the underlying problems of a patient so that timely interventions can be taken for the patient.

\subsubsection{\revise{Mortality Prediction in the MIMIC-III Dataset}}
\label{subsubsec:mortality prediction in mimic-iii dataset-patient level}

\begin{figure}
\centering
\includegraphics[width=0.8\linewidth]{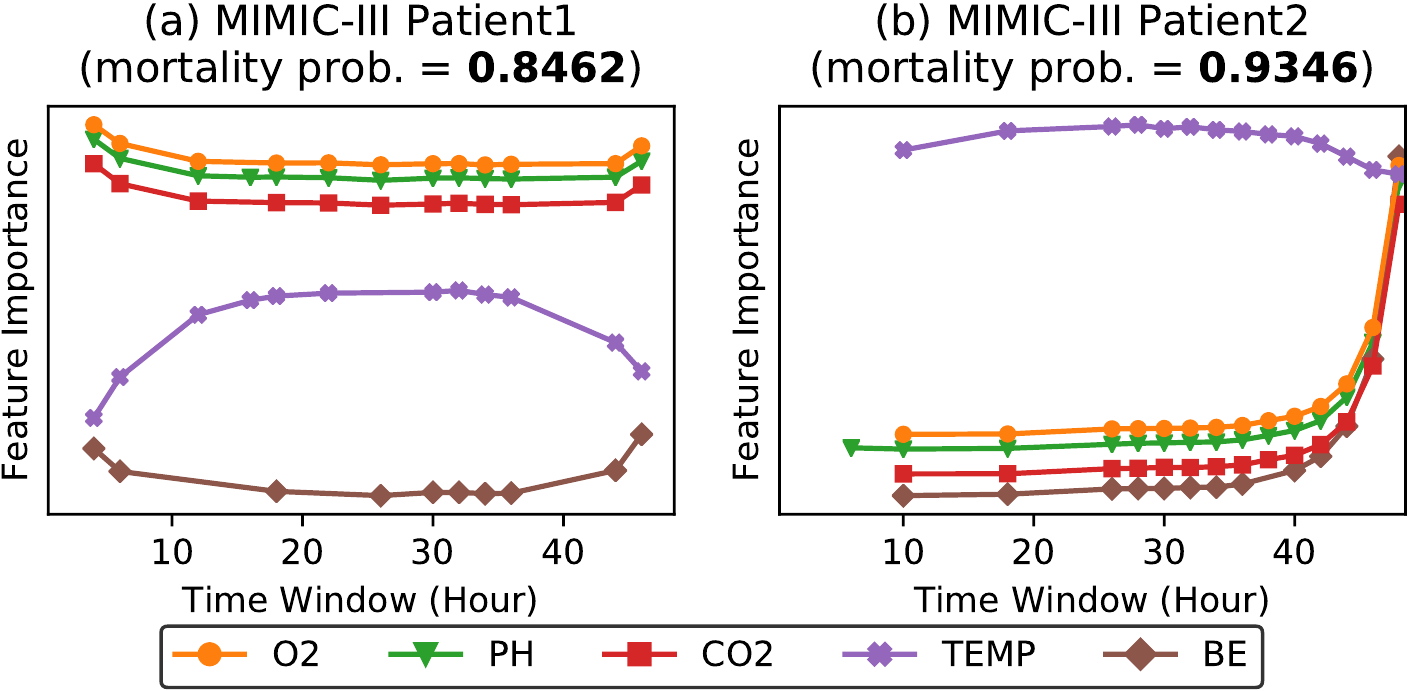}
\caption{\revise{Patient-level interpretation results of \framework
in the MIMIC-III dataset.
}}
\label{fig:mimic patient-level interpretation results}
\end{figure}

\revise{In the in-hospital mortality prediction, we probe into two representative patients who passed away with the Feature Importance changing patterns of five features illustrated in Figure~\ref{fig:mimic patient-level interpretation results}, corresponding to ``Oxygen'' (\Oxygen), ``pH'' (\PH), ``Carbon Dioxide'' (\CarbonDioxide), ``Temperature'' (\TEMP) and ``Base Excess'' (\BE).}

\revise{Among these five, four features \Oxygen, \CarbonDioxide (which in this case of the MIMIC-III dataset, reflects \PCarbonDioxide in blood gas analysis), \PH and \BE are medically closely related in patients' metabolism, respiratory status, and acid-base balance. These are intimately related to major organ functions and illness acuity. For instance, if a patient in the ICU is suffering from inadequate oxygenation or ventilation, which will cause a decrease of \Oxygen or an increase in \PCarbonDioxide respectively, the latter reflects respiratory acidosis. Concurrently, the decreasing \BE reflects worsening metabolic acidosis which in turn suggests inadequate acid-base compensation by the deteriorating kidney function of the patient. The net result is a lower than normal \PH (acidemia). When we investigate the two patients in Figure~\ref{fig:mimic patient-level interpretation results}, we observe that the aforementioned four features tend to exhibit similar Feature Importance changing patterns, possibly due to their similar clinical functionalities in acid-base balance~\cite{breen2001arterial}.}

\revise{Furthermore, \TEMP is medically shown to be highly related to mortality~\cite{bota2004body, lee2012association}. This can be exemplified in Figure~\ref{fig:mimic patient-level interpretation results}, in which both patients have a \TEMP with a relatively large Feature Importance value along with time.}

However, when we compare Patient1 in Figure~\ref{fig:mimic patient-level interpretation results} (a) and Patient2 in Figure~\ref{fig:mimic patient-level interpretation results} (b), we can find that Patient1's mortality is more highly associated with derangements in oxygenation, ventilation, and acid-base derangements; Patient2's mortality seems more associated with extremes of \TEMP which may be the case of severe infection.

With such detailed patient-level analysis of \framework, doctors can better understand the possible terminal processes of a deteriorating patient and
recognize the ones that are more associated with mortality, so that priorities in therapeutic options can be identified in a personalized manner.

\subsection{\revise{Feature-Level Interpretation}}
\label{subsec:feature-level interpretation results}

In this section, we show some feature-level interpretation results in both applications to demonstrate how \framework functions in the feature level, e.g., helps unveil the characteristics of medical features in both applications. Hence, \framework can provide medically meaningful insights contributing to medical research advancement. 
In each application, we use the best-performing checkpoint of the \model model in \framework to plot the distribution
of Feature Importance - Time Window of each feature among all samples.

\newpage
\subsubsection{\revise{AKI Prediction in the NUH-AKI Dataset}}
\label{subsubsec:aki prediction in hospitalx-aki dataset-feature level}

\begin{figure}[t]
\centering
\includegraphics[width=0.8\linewidth]{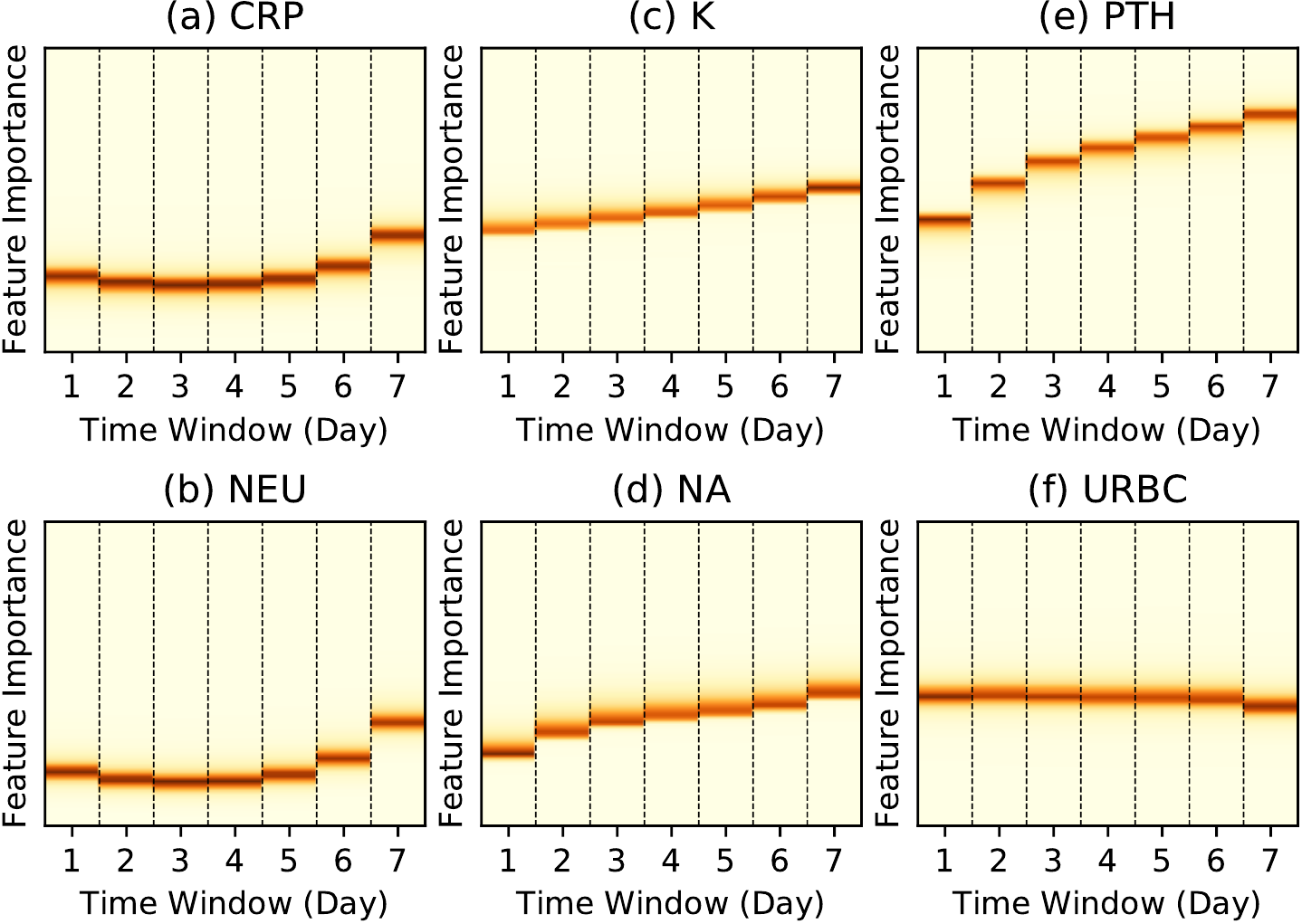}
\caption{Feature-level interpretation results of \framework
in the NUH-AKI dataset.
}
\label{fig:aki feature-level interpretation results}
\end{figure}

\revise{In the NUH-AKI dataset, we illustrate the Feature Importance generated by \framework for six representative laboratory tests in Figure~\ref{fig:aki feature-level interpretation results}, corresponding to ``C-Reactive Protein'' (\CRP), ``Neutrophils'' (\NEU), ``Serum Potassium'' (\K), ``Serum Sodium'' (\NA), ``Parathyroid hormone, Intact'' (\PTH) and ``RBC, Urine'' (\URBC). Interesting patterns from these laboratory tests are discovered, including varying patterns (i.e., \CRP, \NEU, \K, \NA, \PTH) and a stable pattern (i.e., \URBC). Our clinical collaborators have concurred that the observed patterns are validated and proven to be medically meaningful. The detailed explanations for the observed patterns in Figure~\ref{fig:aki feature-level interpretation results} are as follows.}

\vspace{2mm}
\noindent
\powerpoint{Similar patterns discovered in similar features.}
\revise{As shown in Figure~\ref{fig:aki feature-level interpretation results} (a) and (b), \CRP and \NEU tend to share a similar Feature Importance changing pattern along with time. This observation is validated by doctors. According to medical literature and doctors' domain expertise, increasing \CRP suggests worsening systemic inflammatory activity in response to various disease stressors including active infection, myocardial infarction, and cancers~\cite{kim2015c, liuzzo1994prognostic, dolan2017role}, and systemic inflammation may be directly involved in pathogenesis of AKI by altering the kidney microcirculation adversely~\cite{wu2007evidence, lai2016c}.
Likewise, \NEU is the most abundant type of white blood cells in human blood that responds to bacterial infection, cancers, vascular diseases, and the former is an important mediator of inflammation~\cite{mishalian2017diversity, qi2017neutrophil, scapini2014social}.
Due to their similar medical functionality, \CRP and \NEU have a similar response pattern with a sudden increase in the latter part of the time, shown in their Feature Importance changing patterns with time.}

\revise{Similarly, \K and \NA also exhibit similar Feature Importance - Time Window patterns as shown in Figure~\ref{fig:aki feature-level interpretation results} (c) and (d). This is because both \K and \NA-water balance are important fluid and electrolytes in the human body which are vital to cellular metabolism and regulated by the kidneys~\cite{black1969the}. A patient would suffer derangements in \K and \NA balance as kidney function deteriorates~\cite{jung2016electrolyte, gao2019admission}, and these imbalances develop concurrently. Therefore, \K and \NA behave similarly in terms of Feature Importance changing patterns with time.}

\vspace{2mm}
\noindent
\powerpoint{Different patterns indicate different clinical functionalities.} \revise{Based on the analysis above, we can see that compared with \CRP and \NEU, \K and \NA should behave differently in terms of Feature Importance changing pattern along with time as shown in Figure~\ref{fig:aki feature-level interpretation results}, due to their different clinical functionalities.}

\revise{Besides, \PTH is the primary regulator of systemic calcium and phosphate homeostasis in humans and regulates transporters to increase excretion of filtered phosphate in the kidneys~\cite{pfister1997parathyroid}.
The hypocalcemia and hyperphosphatemia that develop with AKI may up-regulate \PTH activity. Skeletal resistance to \PTH is also observed in kidney failure~\cite{somerville1978resistance}. These observations explain the elevated \PTH observed in patients with AKI~\cite{leaf2012fgf, vijayan2015relationship}. The closer in time the measured \PTH is to Prediction Window of AKI prediction, the more significant its feature importance will be. As shown in Figure~\ref{fig:aki feature-level interpretation results} (e), \PTH shows a high influence on the AKI risk, denoted as a relatively high Feature Importance which increases in significance along with time.}

\revise{Furthermore, \URBC (Figure~\ref{fig:aki feature-level interpretation results} (f)) exerts a relatively stable influence on the AKI risk in terms of Feature Importance with time. The presence of \URBC infers hematuria (blood in urine). Hematuria is commonly observed in kidney (glomerular) diseases and associated kidney dysfunction~\cite{yuste2016haematuria}. Hematuria may suggest a compromise in the glomerular filtration barrier, and its presence has been shown to be strongly associated with kidney disease progression over time~\cite{moreno2016haematuria}. Therefore, \URBC may exhibit a stable Feature Importance on the AKI prediction.}

\subsubsection{\revise{Mortality Prediction in the MIMIC-III Dataset}}
\label{subsubsec:mortality prediction in mimic-iii dataset-feature level}

\begin{figure}[t]
\centering
\includegraphics[width=0.8\linewidth]{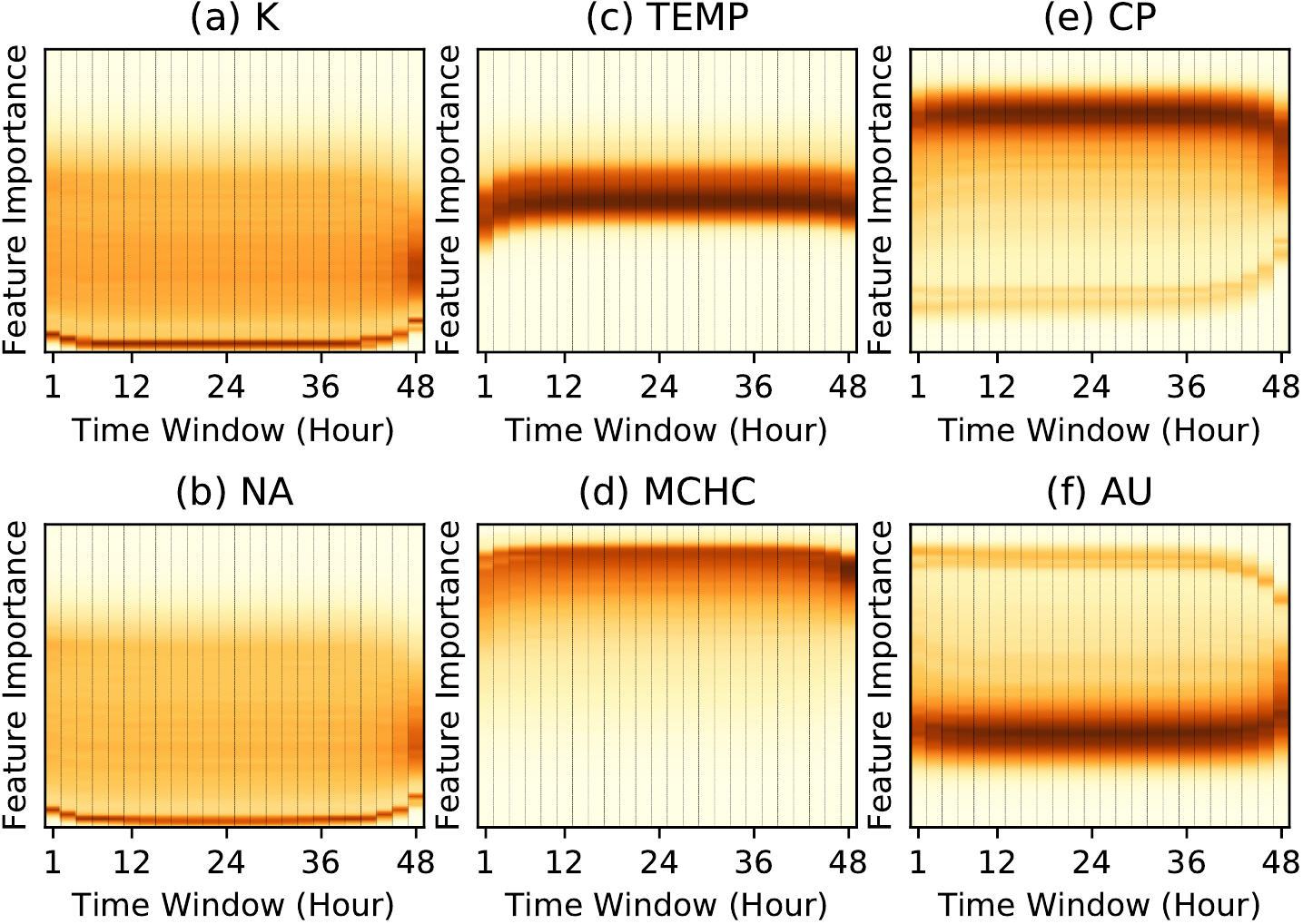}
\caption{Feature-level interpretation results of \framework
in the MIMIC-III dataset.
}
\label{fig:mimic feature-level interpretation results}
\end{figure}

\revise{In the MIMIC-III dataset, we illustrate the Feature Importance changing patterns along with time for six representative features in Figure~\ref{fig:mimic feature-level interpretation results}, corresponding to ``Serum Potassium'' (\K), ``Serum Sodium'' (\NA), ``Temperature'' (\TEMP), ``Mean Corpuscular Hemoglobin Concentration'' (\MCHC), ``Cholesterol, Pleural'' (\CP) and ``Amylase, Urine'' (\AU). We examine the changing patterns and summarize our findings (which have been validated by our clinical collaborators to be medically meaningful) with the detailed medical correlation.}

\vspace{2mm}
\noindent
\powerpoint{\revise{Low Feature Importance detected for common features which are not generally highly related to mortality.}} \revise{For \K and \NA as illustrated in Figure~\ref{fig:mimic feature-level interpretation results} (a) and (b), they exhibit the following Feature Importance - Time Window patterns: (i) a flat curve with low Feature Importance values with some fluctuations; (ii) a noisy area with Feature Importance values dispersing over the whole area. We suggest that this phenomenon occurs due to the characteristics of \K and \NA balance in critically ill patients. These electrolyte disorders are very common in critically ill patients~\cite{lee2010fluid}; minor abnormalities may be too general to exert significant causal effect and hence, not be highly related to mortality. This helps explain the flat curve (i) in the figure. We note however that there are certain cases when such common features are related to mortality; patients would have varying severity of \K and \NA imbalance that persist due to the illness acuity, poor nutritional intake in worsening disease, intravenous fluids administered in huge quantity, or loss of body fluids in unique clinical situations with high gastrointestinal losses~\cite{gao2019admission, winata2019intravenous}. These may lead to various changing patterns of Feature Importance with time, which corresponds to the noisy area (ii) in the figure.}

\vspace{2mm}
\noindent
\powerpoint{\revise{High Feature Importance detected for common features that are generally highly related to mortality.}} \revise{As shown in Figure~\ref{fig:mimic feature-level interpretation results} (c), \TEMP has a relatively large and stable Feature Importance with time, which means that \TEMP is relatively highly related to mortality. This observation is medically plausible. Extremes of fever and body temperature have prognostic implications in sepsis and are among the key criteria in defining a systemic inflammatory response syndrome~\cite{bone1992definitions}. The alterations in \TEMP could be infection-related or a host's response to inflammatory stress of non-infectious origin~\cite{o2008guidelines, hawksworth2009new}. High fever could predispose to cardiac arrhythmias, increased oxygen demand, seizures and brain injury in patients and this portends adverse outcomes~\cite{manthous1995effect, bernard2002treatment}. Both hyperthermia and hypothermia, and even pharmacological temperature lowering, are associated with higher mortality in critically ill patients~\cite{bota2004body, lee2012association}.}

\revise{Similar to \TEMP, \MCHC's Feature Importance remains large with time shown in Figure~\ref{fig:mimic feature-level interpretation results} (d); it poses a larger influence on mortality than other appearing features. Low \MCHC indicates low hemoglobin concentration per red blood cell in the circulation, and may imply lower oxygen-carrying capacity by blood cells to the tissues. This may explain the observation that lower \MCHC is associated with mortality in patients with myocardial infarction in ICU~\cite{huang2016lower}. In addition, patients with sepsis and critical illness often develop acidemia or hypophosphatemia, which in turn alters hemoglobin-oxygen affinity and reduces oxygen release to tissues~\cite{watkins1974left}. Therefore, \MCHC's effect on mortality might relate to downstream effects on end-organ malperfusion.}

\vspace{2mm}
\noindent
\powerpoint{\revise{Same feature's diverging patterns indicate different patient clusters.}} \revise{As for \CP (Figure~\ref{fig:mimic feature-level interpretation results} (e)) and \AU (Figure~\ref{fig:mimic feature-level interpretation results} (f)), we observe that their corresponding Feature Importance changing patterns with time exhibit an apparent diverging phenomenon. We suppose such diverging patterns indicate the clinical functionality of both features in helping divide patients into different clusters.}

\revise{Specifically, \CP is examined to be of supportive clinical value in differentiating two types of pleural effusion: exudative (higher CP) and transudative~\cite{hamal2013pleural}. Exudative pleural effusions either follow acute or chronic lung infection, or lung cancer; these conditions may relate to more adverse patient outcomes and hence the association with ICU mortality. On the other hand, low \CP and transudative effusion may follow severe volume overload in the setting of critical illness and organ injury, and there are multiple studies demonstrating an association with fluid overload and increased ICU mortality~\cite{bouchard2009fluid, payen2008positive}.
}

\revise{Similarly, the Feature Importance - Time Window pattern of \AU is novel and interesting. It also diverges into two types, indicating the presence of different patient clusters. \AU level is correlated to serum amylase level~\cite{terui2013urinary}. Serum amylase in turn is elevated in severe clinical diseases including acute pancreatitis~\cite{keim1998comparison}, as well as non-pancreatic abdominal organ injury in trauma~\cite{kumar2012evaluation}, and also elevated just with kidney dysfunction due to reduced clearance~\cite{seno1995serum}. In the latter, low \AU may occur despite raised serum levels, which may explain the diverging patterns illustrated in Figure~\ref{fig:mimic feature-level interpretation results} (f).}

\subsection{\rev{\framework for Financial Analytics}}
\label{subsec:other applications}

\rev{
We have thus far focused on evaluating \framework in healthcare analytics.
In this section, we will move on to demonstrate how to employ \framework in other high stakes applications exemplified with financial analytics first,
whose performance can be greatly improved with automated financial analytic algorithms~\cite{Robot_Analysts}.
Among various financial analytics, stock index prediction is of critical importance for the investment and trading of financial professionals~\cite{krollner2010financial}.
}

\begin{figure}
	\centering
	\includegraphics[width=0.8\linewidth]{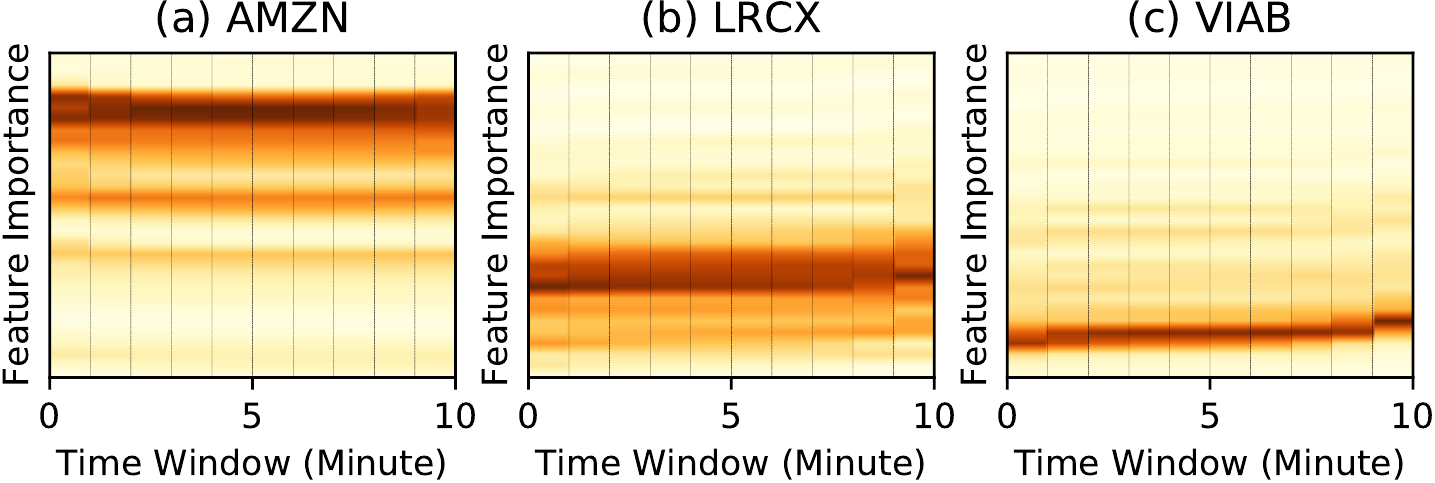}
	\caption{\rev{Feature-level interpretation results of \framework in the NASDAQ100 dataset.}}
	\label{fig:nasdaq results}
\end{figure}

\rev{
We evaluate \framework in the real-world stock index prediction of NASDAQ-100 Index.
Specifically, we use the NASDAQ100 dataset~\cite{qin2017dual}, which collects the stock prices of 81 major corporations in NASDAQ-100 and the NASDAQ-100 Index values every minute from July 26, 2016 to December 22, 2016.
This prediction is a regression task for the current NASDAQ-100 Index with recent stock prices of the 81 constituent corporations and the NASDAQ-100 Index values.
Therefore, the index value of each minute is a prediction target and thus corresponds to one sample~\cite{qin2017dual}.
In the experiment, the time window is set to one minute and Feature Window $10$ minutes.

In Figure~\ref{fig:nasdaq results}, we show the feature-level interpretation results of three representative stocks: Amazon.com, Inc. (AMZN), Lam Research Corporation (LRCX), and Viacom Inc. (VIAB) which are a top-ranking, mid-ranking and bottom-ranking stock respectively in NASDAQ-100 Index.
Figure~\ref{fig:nasdaq results} illustrates the feature-level interpretation of all prediction samples and thus, the dispersion along y-axis of each feature indicates the variability of the corresponding feature importance in the observed time span.

We can observe in Figure~\ref{fig:nasdaq results} that Feature Importance of the three stocks is quite stable over different time windows.
This is because the prediction is made based on a $10$-minute Feature Window; therefore, the stability of stock prices is anticipated within such a short period of time.

Further, we can notice that the three stocks exhibit different Feature Importance changing patterns:
\begin{itemize}[leftmargin=*]
    \setlength\itemsep{3mm}
    \item
    AMZN~\cite{AMZN}, a top-ranking stock, has a high but fluctuating Feature Importance in the prediction.
    The findings demonstrate that AMZN has a significant influence on NASDAQ-100 Index, and the variance of such influence (either increase or decrease) is large, which reveals that AMZN is an important indicator of the index.
    
    \item LRCX~\cite{LRCX}, a mid-ranking stock, has a medium Feature Importance across time, with moderate fluctuations among predictions.
    This shows that although not predominant, such stocks are important constituents of NASDAQ-100 Index.
    Further, the dispersion of Feature Importance among different samples varies mildly in different time periods, which shows that LRCX is a valuable indicator of the index.
    
    \item VIAB, a bottom-ranking stock, shows a consistently low Feature Importance changing pattern with minor fluctuations.
    The interpretation results reveal that VIAB has a small influence on NASDAQ-100 Index.
    This is corroborated by a later announcement in December, 2017 by NASDAQ that VIAB is removed from the index after re-ranking~\cite{VIAB_News}.
\end{itemize}
 
In a nutshell, the interpretation results can reveal not only the importance but also the variability of the importance of each stock.
The availability of such information is critical in the decision making for financial professionals in investment and risk management~\cite{Risk_Management} during the management of their portfolios.
The experiments confirm the capability of \framework in providing insights for the high stakes financial application.
}

\subsection{\rev{\framework for Temperature Forecasting}}

In this section, we demonstrate how to employ \framework in temperature forecasting, a critical application that targets at reducing power consumption and hence, improving energy efficiency~\cite{zamora2014line}.

\begin{figure}
	\centering
	\includegraphics[width=0.65\linewidth]{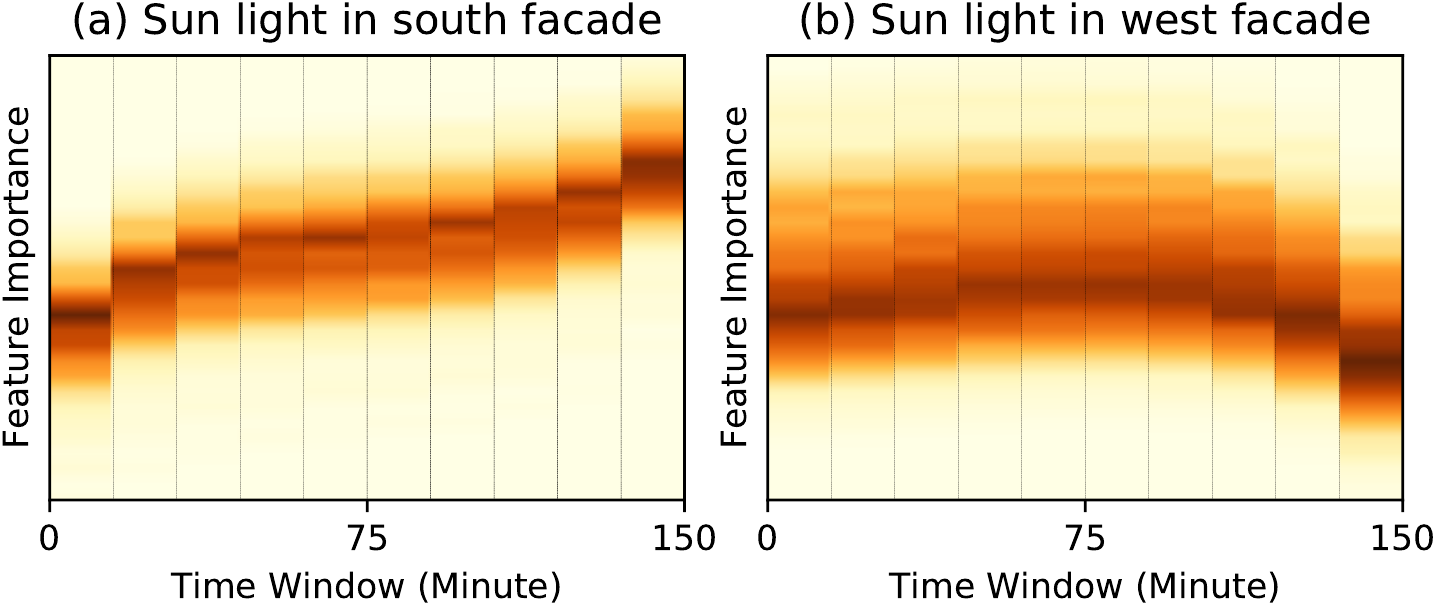}
	\caption{\rev{Feature-level interpretation results of \framework in the SML2010 dataset.}}
	\label{fig:sml2010 results}
\end{figure}

\rev{
	We evaluate \framework in the indoor temperature forecasting application. To be specific, we use the SML2010 dataset~\cite{zamora2014line}, a public dataset from UCI Machine Learning Repository~\cite{Dua:2019} which is collected from SMLsystem from March 2012 to May 2012. In this application, we aim to predict the current indoor temperature as a regression task, given $16$ time-series sensor data sources as input features. In the experiment, the time window is set to $15$ minutes and Feature Window $150$ minutes.
	
	In Figure~\ref{fig:sml2010 results}, we illustrate the feature-level interpretation results of \framework for two representative features: sun light in south facade ($SL_{south}$), and sun light in west facade ($SL_{west}$), as sun light intensity is apparently one of the key influential factors to the indoor temperature.
	According to the interpretation results provided by \framework in Figure~\ref{fig:sml2010 results}, we can see that $SL_{south}$ and $SL_{west}$ are both important, but exhibit different characteristics in terms of Feature Importance changing patterns.
	
\begin{itemize}[leftmargin=*]
	\setlength\itemsep{3mm}
	\item $SL_{south}$'s rising Feature Importance across time. 
	We note that this SML2010 dataset is collected from March to May at CEU-UCH in Valencia, Spain~\cite{zamora2014line}, i.e., during the spring in the mid-latitude region where the temperature differs much between day and night, and the sun shines on the south facade mostly in the daytime, yet shines on the west facade in the evening.
	Therefore, $SL_{south}$ can represent the real-time sun light intensity. The nearer to prediction in time, the larger influence $SL_{south}$ can pose to the indoor temperature. This agrees with the rising Feature Importance of $SL_{south}$  along with time calculated by \framework.
	
	\item $SL_{west}$'s stable Feature Importance over time. 
	Different from the south facade, the sun shines on the west facade only in the evening when it is relatively dark. This causes $SL_{west}$ to serve as a relatively stable indicator of outdoor darkness (e.g., daytime vs. night, sunny vs. cloudy). Hence, $SL_{west}$ exhibits a relatively stable Feature Importance over time, with a slight decrease approaching the prediction time. This decrease appears due to the fact that in the time windows near prediction, other features that can represent the real-time sun light intensity such as $SL_{south}$ are relatively more important. These findings are well reflected by the interpretation results from \framework.

\end{itemize}
	
	Based on the experimental results and analysis above, \framework is shown to help unveil the characteristics of features in this indoor temperature forecasting application and hence, provide meaningful information to the corresponding domain practitioners. As a result, we confirm the applicability of \framework in this critical temperature forecasting application.}

\section{Conclusions}
\label{sec:conclusion}
Interpretability has been recognized to play an essential role in designing \rev{analytic models in high stakes applications such as healthcare analytics}. Feature importance is one common way to interpret the predictions of analytic models. \revise{In this paper, we propose to capture the feature importance in two aspects: the time-invariant and the time-variant feature importance, respectively reflecting the overall influence of the feature shared across time and the time-related influence which may vary along with time.}
We devise \model to model the time-invariant feature importance via feature-wise transformation, and the time-variant feature importance via self-attention. 
With \model as the core component, 
we propose a framework \framework to \rev{provide accurate and interpretable clinical decision support to doctors and insightful advice to practitioners of other high stakes applications as well.}
\rev{We first evaluate the effectiveness of \framework in healthcare analytics by conducting extensive experiments} over the NUH-AKI dataset and the MIMIC-III dataset for AKI prediction and mortality prediction.
The results show that \framework is able to provide more accurate predictions than all the baselines in both datasets.
Further, the interpretation results have also been  validated to be medically meaningful by the clinicians.
\rev{We also evaluate \framework in a financial application and a temperature forecasting application to demonstrate its applicability in other high stakes applications.}

\section{Acknowledgments}
This research is supported by the National Research Foundation Singapore under its AI Singapore Programme (Award Number: AISG-GC-2019-002). Besides, this research is also supported by AI Singapore 100 Experiments (Award Number: AISG-100E-2018-007).

\clearpage
\begin{spacing}{0.99}
\bibliographystyle{abbrv}
\bibliography{main_full_arxiv}
\end{spacing}

\end{document}